\documentclass[
aip,
amsmath,amssymb,
preprint,
superscriptaddress,
]{revtex4-2}

\usepackage{graphicx}
\usepackage{dcolumn}
\usepackage{bm}

\usepackage{units}
\usepackage[dvipsnames]{xcolor}
\usepackage{comment}

\graphicspath{ {figures/} }	

\usepackage[acronym]{glossaries}
	\newacronym{FEM}{FEM}{Finite Element Method}
	\newacronym{MEMS}{MEMS}{Micro-Electro-Mechanical Systems}
	\newacronym{2D}{2D}{two-dimensional}
	\newacronym{3D}{3D}{three-dimensional}
	\newacronym{BSOI}{BSOI}{Bonded Silicon on Insulator}
	\newacronym{CMUT}{CMUT}{Capacitive Micromachined Ultrasonic Transducers}

\begin{document}


\title{
Coulomb actuated microbeams: A Chebyshev-Edgeworth approach to highly efficient lumped parameter models}

\author{Hermann A. G. Schenk}%
\email{hermann.schenk@arioso-systems.com}
\affiliation{%
Arioso Systems GmbH, Dresden, 01109 Germany}%
\author{Anton Melnikov}
\author{Franziska Wall}
\affiliation{%
Fraunhofer Institute for Photonic Microsystems IPMS, 01109 Dresden, Germany}
\author{Matthieu Gaudet}
\affiliation{%
Universität Oldenburg, Department für Informatik - AMiR, 26111 Oldenburg, Germany}
\author{Michael Stolz}
\affiliation{%
Fraunhofer Institute for Photonic Microsystems IPMS, 01109 Dresden, Germany}
\affiliation{%
Brandenburg University of Technology Cottbus-Senftenberg, 03046 Cottbus, Germany}
\author{David Schuffenhauer}
\author{Bert Kaiser}
\affiliation{%
Fraunhofer Institute for Photonic Microsystems IPMS, 01109 Dresden, Germany}

\date{\today}

\begin{abstract}
In a  previous   publication   we   demonstrated   that the  stable  and  unstable  equilibrium states  of  prismatic Coulomb actuated Euler-Bernoulli micro-beams, clamped at both ends, can successfully  be simulated  combining  finite element analysis (FEM) with continuation methods.  Simulation results were  experimentally  scrutinised  by  combining  direct optical  observations  with  a  modal  analysis  regarding Euler-Bernoulli eigenmodes. Experiment and simulation  revealed  convincing  evidence for the possibility of modelling the physics of such a micro-beam by means of lumped parameter models involving only a single degree of freedom, the Euler-Bernoulli zero mode. In this paper we present the corresponding analytical single degree of freedom lumped parameter model (LPM). This comprehensive model demonstrates the impact of the beam bending on the nature of the Coulomb singularity, allows for an easy and accurate computation of the pull-in voltage in the presence of stress stiffening and is apt for efficient frequency response computations. Our method to derive the zero-mode LPM is based on a  Chebyshev-Edgeworth type method as is common in analytical probability theory. While used here for a very particular purpose, this novel approach to non-linear dynamic systems has a much broader scope. It is apt to analyse different boundary conditions, electrostatic fringe field corrections and squeeze film damping, to name a few applications.

\end{abstract}

\maketitle



\section{Introduction}
%
%
%
%

Coulomb-actuated microbeams play a crucial role in many \gls{MEMS} applications \cite{Senturia.2002, Leondes.2006, Hsu.2008}. They enable actuation using electrostatic forces and capacitive sensing, give rise to pioneering applications in medicine\cite{Saliterman.2006}, communications\cite{Lucyszyn.2004, Li.2018}, sensing\cite{Leondes.2006, Coppa.2007}, and consumer products\cite{S.Finkbeiner.2013, Zou.2014, Verdot.2016, Shahosseini.2013, Kaiser.}.
To meet the needs of recent developments, such as 5G Internet of Things (5G-IoT) \cite{Li.2018}, augmented reality \cite{Kim.2009}, and Green ICT (information and communications technology) \cite{Bianzino.2010, Worthington.2017}, a system level consideration of a high number of electro-mechanical components is necessary. This is only possible, if accurate and highly efficient lumped parameter models of the components are available.

In this paper we systematically derive a single degree of freedom lumped-parameter model (LPM), describing the physics of prismatic clamped-clamped Coulomb actuated micro-beams with high precision as compared to FEM simulations and in line with experimental findings.  
In a previous publication, Melnikov \textit{et al.} \cite{Melnikov.2021} demonstrated that the stable and unstable states of prismatic Coulomb actuated Euler-Bernoulli micro-beams, clamped at both ends, can be successfully simulated combining FEM with arc-length solvers. The resulting model predictions were experimentally scrutinised by combining direct optical observations with a modal analysis regarding Euler-Bernoulli eigenmodes. Both approaches revealed convincing evidence for an almost perfect congruence of the respective bending profile and the shape of the lowest Euler-Bernoulli eigenmode (the zero-mode).
It was shown that this is true for the entire applicable voltage range within very small error margins. The observation suggests the possibility to model the physics of such a micro-beam by means of a lumped parameter model involving only a single degree of freedom, amenable to direct physical interpretation.

Studies analytically deriving lumped parameter models, e. g.  Nayfeh, Younis, and Rahman 
\cite{Younis.2002, EihabMAbdelRahman.2002, M.I.Younis.2003, M.I.Younis.2003b, Nayfeh.2005, Nayfeh.2007, Younis.2011}
typically begin with the non-linear Euler-Bernoulli  beam equation for the bending profile $w(\xi,\tau)$. In its dimensionless form used by Nayfeh \textit{et al.} \cite{Nayfeh.2007} this equation reads
\begin{equation}\label{eq:full_beam1}
\begin{aligned}
    &\frac{\partial^2 w}{\partial \tau^2} + c \frac{\partial w}{\partial \tau} + \frac{\partial^4 w}{\partial \xi^4}
    =\\
    &= ( \gamma[w] + N) \frac{\partial^2 w}{\partial \xi^2} + \alpha_2 \frac{v(\tau)^2}{(1-w)^2}.
\end{aligned}
\end{equation}
Here $\xi$ and $\tau$ are the dimensionless beam coordinate and the dimensionless time. The dynamic damping coefficient is denoted by  $c$. The geometry dependent parameters $\alpha_1$ and $\alpha_2$ are given in Eq.~\eqref{eq:LPM} below. $N$ is an dimensionless axial stress and $v(\tau)$ is the dimensionless drive voltage. The beam is assumed to be clamped at $\xi=-\frac{1}{2}$  and at $\xi=+\frac{1}{2}$, where the usual clamped-clamped boundary conditions Supplementary Eq.~(S7) apply. The non-local functional $\gamma[w]$ models the stress stiffening of the clamped-clamped beam,
\begin{equation}
  \gamma[w] = \alpha_1 \int_{- \frac{1}{2}}^{+ \frac{1}{2}} \left( \frac{\partial w}{\partial \xi} \right)^2 \mathrm{d}\xi \ .
\end{equation}
Nayfeh \textit{et al.} expand the bending profile with respect to a complete ortho-normal Hilbert space base $\psi_{n}(\xi)$, 
\begin{equation}\label{eq:eigenmode_expansion}
    w(\xi,\tau) = \sum_{n=0}^{\infty}\ \hat{w}_{n}(\tau) \ \psi_{n}(\xi) \ .
\end{equation}
Upon insertion into Eq.~\eqref{eq:full_beam1}, the partial differential equation Eq.~\eqref{eq:full_beam1} is converted into an infinite set of coupled nonlinear ordinary differential equations of the form \cite{Younis.2002, EihabMAbdelRahman.2002, M.I.Younis.2003, M.I.Younis.2003b, Nayfeh.2005, Nayfeh.2007, Younis.2011}
\begin{equation}\label{eq:modal_ODEs}
    \frac{\partial^2  \hat{w}_{n}}{\partial \tau^2} + c \frac{\partial  \hat{w}_{n}}{\partial \tau} +  \sum_{m=0}^{m=\infty} k_{n,m}[w] \   \hat{w}_{m}
    = \alpha_2 v^2 F_n[\tau,w].
\end{equation}
Unlike Nayfeh \textit{et al.}, we select ${\{ \lambda_n, \psi_n(\xi)}\}_{n\in \mathbb{N}}$ to be the Euler-Bernoulli eigen system and can therefore be a little more specific,
\begin{equation}
     k_{n,m}[w] = \lambda_n \delta_{n,m} + (\gamma [w] + N )\, \chi_{n,m} \, ,
\end{equation}
\begin{equation}
  \chi_{n,m} = \int_{- \frac{1}{2}}^{+ \frac{1}{2}} 
  \frac{\partial\psi_{n}}{\partial\xi} 
  \frac{\partial\psi_{m}}{\partial\xi} 
  \mathrm{d}\xi\ .
\end{equation}
The challenge with this approach however is that the resulting stiffness matrix $k_{n,m}[w]$ and the force components $F_{n}[\tau,w]$ are rather intricate,  non-linear, singular and time dependent functionals of the entire infinite set of the coefficient functions $ \lbrace \hat{w}_{n}(\tau) \rbrace_{n\in  \mathbb{N}} $ : 
\begin{equation}
\begin{aligned}
     k_{n,m}[w] &= k_{n,m}[ \hat{w}_{0}(\tau), ...,  \hat{w}_{n}(\tau), ...], \\
    \\
     F_n[\tau,w] &= F_n[\tau, \hat{w}_{0}(\tau), ...,  \hat{w}_{n}(\tau), ...].
\end{aligned}
\end{equation}
This circumstance makes it in general very challenging to obtain any elucidating results from Eq.~\eqref{eq:modal_ODEs}.
As can be see from literature, the complexity of the functionals $k_{n,m}[w]$ and $F_n[\tau,w]$ leads to a tedious computational task, even after introducing well considered simplifications, e.g. see Younis \textit{et al.} \cite{M.I.Younis.2003b}.
The resulting computations seem neither more attractive than direct numerical methods, nor is the need for the number of degrees of freedom, required to obtain satisfactory accuracy, amenable to direct physical interpretation. In fact the number of modes required in Nayfeh's \textit{et al.} approach turns out to be an artefact, essentially reflecting their comparatively straight forward attempt to technically cope with the singular nature of the Coulomb force, as we will see.      

The picture substantially changes however with the observation of Melnikov \textit{et al.} \cite{Melnikov.2021} that the lowest Euler-Bernoulli eigenmode $\psi_0(\xi)$ is by far dominating the physics of  Coulomb actuated prismatic clamped-clamped micro-beams in practical applications. 
This observation implies that the use of higher modes in a LPM for a prismatic Euler-Bernoulli beam is hardly justified, unless higher kinetic energies are involved. Due to the large spectral distance, typically a multiple of the elastic energy corresponding to the considered deflection of the zero-mode  is required for significant effects involving higher modes.

The  observation of Melnikov \textit{et al.} \cite{Melnikov.2021} essentially allows to reduce the Eq.~\eqref{eq:eigenmode_expansion} to the single term

\begin{equation}\label{eq:zero_mode_ansatz}
   \begin{aligned}
    w(\xi,\tau) &\approx z(\tau) \ \frac{\psi_{0}(\xi)}{\psi_{0}(0)}, \quad 0 \le z(\tau) \le 1 \ , \\
    \psi_0(\xi) &= 
    \frac{\cosh(\beta_0 \xi)}{\cosh(\beta_0/2)} 
    -\frac{\cos(\beta_0 \xi)}{\cos(\beta_0/2)} \ .
    \end{aligned}
\end{equation}
Here $\beta_0$ is the smallest solution to the equation
\begin{equation}\label{eq:beta_definition}
    0 = \tanh (\beta/2) + \tan(\beta /2) \ . 
\end{equation}
In zero-mode approximation  Eq.~\eqref{eq:modal_ODEs} simplifies to the quite handy form 
\begin{equation}\label{eq:basic_LPM}
    \frac{\partial^2}{\partial \tau^2} z + c \frac{\partial}{\partial \tau} z  
    + k_0 \, z + \kappa \, z^3 \,
    = u^2 f_0(z) \ , \\
\end{equation}
The parameters $\kappa$,  $k_0$ and u are defined as 
\begin{equation}
    \begin{aligned}
        \kappa &= \alpha_1 \left( \frac{\chi_0}{\psi_0(0)} \right)^2 , \\
           k_0 &=\lambda_0 + N \chi_0 \quad , \quad
                u = \psi_0(0) \sqrt{\alpha_2} \, v  \, , \\
    \end{aligned}
\end{equation}
\begin{equation}\label{eq:chi0}
    \chi_0 = \int_{- \frac{1}{2}}^{+ \frac{1}{2}} 
            \left(\frac{\partial\psi_{0}}{\partial\xi}\right)^2
            \mathrm{d}\xi
\end{equation}

and the force term is
\begin{equation}\label{eq:Coulomb_f0_def}
    f_0(z) = \int_{-\frac{1}{2}}^{+ \frac{1}{2}} 
    \frac
    {\frac{\psi_0(\xi)}{\psi_0(0)}}
    {\left( 1 - z \, \frac{\psi_0(\xi)}{\psi_0(0)} \right)  ^2 
    } \, \mathrm{d}\xi \  .\\
\end{equation}
The remaining key challenge, and the prime topic of this paper, is of course evaluating the Coulomb integral $f_0(z)$. This requires a non-pertubative treatment of the Coulomb singularity. The ad-hoc approach of Younis \textit{et al.} \cite{M.I.Younis.2003b} essentially creates an artificial need for higher modes and therefore enforces dealing with a coupled system of non-linear ordinary differential equations (ODE). This is far from satisfactory.
It is the purpose of this paper to demonstrate, in contrast, that the physics of a Coulomb actuated prismatic Euler-Bernoulli is contained in the single ODE Eq.~\eqref{eq:basic_LPM} to an extend sufficient for most practical purposes in MEMS technology. To this end we devise a non-pertubative strategy of dealing with the Coulomb integral, based on a  Chebyshev-Edgeworth type expansion\cite{Tchebycheff.1890, Edgeworth.1905}. As a result we arrive at a highly accurate analytical expression for  $f_0(z)$. Finally, the application of our zero-mode LPM Eq.~\eqref{eq:LPM} to the simulation results and experimental findings of Melnikov \textit{et al.} \cite{Melnikov.2021}, reveal a very good agreement.

\section{Results}

\subsection{Chebyshev's argument}

 
Our evaluation the of integral $f_0[z]$ begins with the series representation
\begin{equation}\label{eq:f0_In_series}
    f_0(z) = \sum_{n = 1}^{\infty} n \, I_n \, z^{n-1} \ ,
\end{equation}

where the integrals $I_n$ are defined as 

\begin{equation}\label{eq:In_integral_1}
    \begin{aligned}
    I_n =  \int_{- \frac{1}{2}}^{+ \frac{1}{2}} \left( \frac{\psi_{0}(\xi)}{\psi_{0}(0)} \right)^n \, \mathrm{d}\xi  \ .
    \end{aligned}
\end{equation}

Note that because $\lvert I_n \rvert < 1$ we can infer by means of the Cauchy-Hadamard theorem that the series Eq.~\eqref{eq:f0_In_series} is absolutely convergent in the open disc $\lvert z \rvert < 1$, as required for our purposes. The integrals $I_n$ can be cast into the form,
\begin{equation}\label{eq:In_integral_2}
  I_n = \frac{\sqrt{2 \pi}}{\sigma \sqrt{n}} \int_{-\frac{1}{2}\sigma \sqrt{n}}^{+\frac{1}{2}\sigma \sqrt{n}} \Phi_n(\xi) \, \mathrm{d}\xi \, ,
\end{equation}
 where $\Phi_n(\xi)$ is defined as 
\begin{equation}\label{eq:phi_n}
    \begin{aligned}
         \Phi_n(\xi) &= \frac{1}{\sqrt{2 \pi}}
            \left( 
                \frac{1}{\psi_0(0)}\psi_0 
                \left( \frac{\xi}{\sigma \sqrt{n}} \right)
            \right)^n\ ,\\
          \sigma^2 &= -\frac{\psi_0^{(2)}(0)}{\psi_0(0)} \, .\\
    \end{aligned}
\end{equation}
Our strategy now is to evaluate the limiting function 
$\Phi_\infty(\xi)$ 
of the sequence  
$\{ \Phi_n(\xi) \}_{n \in \mathbb{N}} $  
and subsequently to expand  
$\Phi_n(\xi)$ 
around 
$\mathit{n}=\infty$ 
with respect to 
$\mathit{n}^{-1}$. 
This allows us to explicitly perform the integration 
Eq.~\eqref{eq:In_integral_2}. 
As a result we can perform the summation 
Eq.~\eqref{eq:f0_In_series}.
This way we arrive at the targeted formula for 
$f_0(z)$. 

\begin{figure}[hbtp]
	\centering
	\includegraphics[height = 6.5cm]{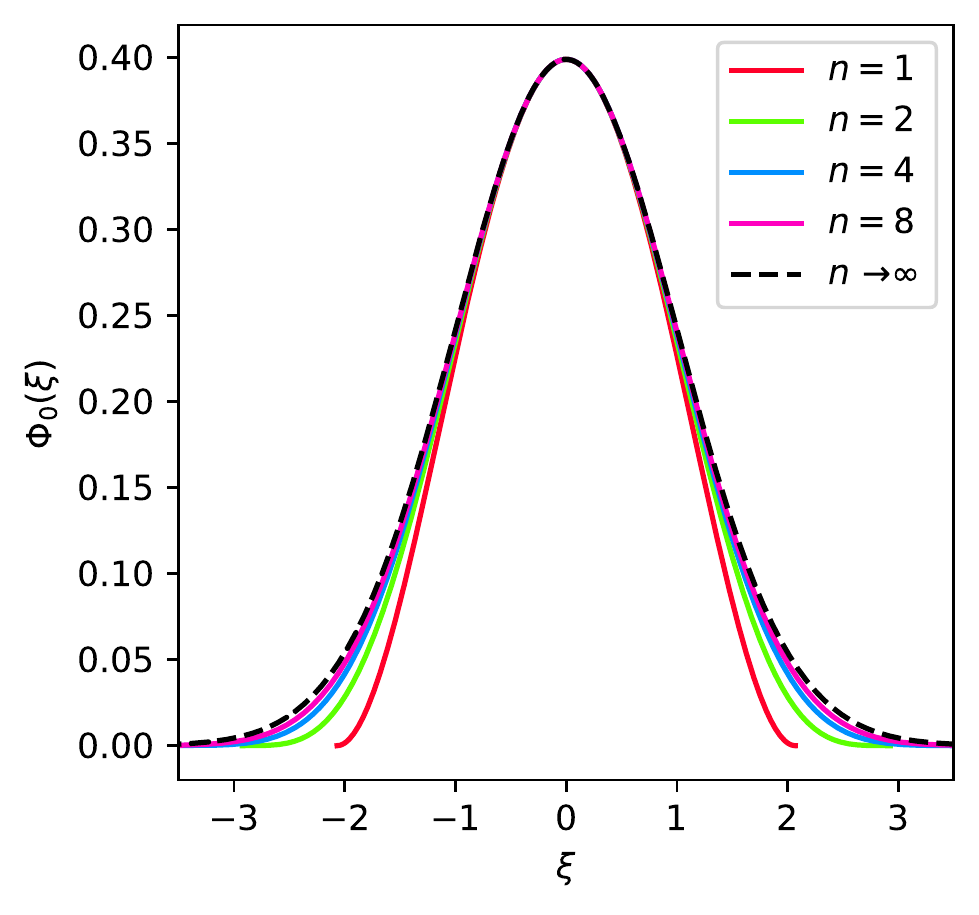}
	\caption{ $\Phi_n(\xi)$ is compared for $n = 1,2,4,8 $ (colored lines) to the Gauss bell curve (dashed black line) illustrating the rapid convergence according to Eq.~\eqref{eq:normal_distribution}. }
	\label{fig:normal_distribution}
\end{figure}

The crucial observation regarding the limiting function 
$\Phi_\infty(\xi)$ 
is that the sequence  
$\{\Phi_n(\xi)\}_{n \in  \mathbb{N}} $ 
uniformly converges to the shape of the Gauss bell curve, 
\begin{equation}\label{eq:normal_distribution}
     \lim_{n\to\infty} \Phi_n(\xi) =
     \frac{1}{\sqrt{2 \pi}} \exp \left( -\frac{\xi^2}{2}\right) \, .
\end{equation}
This important fact is illustrated in Fig.~\ref{fig:normal_distribution}. To motivate how this comes about, we remind the reader of Euler's elementary definition of the exponential function, presented here in a form suitable for our purposes,
\begin{equation}
     \lim_{n\to\infty} \frac{1}{\sqrt{2 \pi}} \left( 1-\frac{\xi^2}{2 n}  \right)^n =
     \frac{1}{\sqrt{2\pi}} \exp \left( -\frac{\xi^2}{2}\right).
     \label{eq:elementary_exp}
\end{equation}
The quite perplexing idea that  Eq.~\eqref{eq:elementary_exp} holds for a much broader class of functions, inserted into its left hand side, dates back to the ground breaking contributions of P.L.~Chebyshev to the field of analytical probability theory.\cite{Tchebycheff.1890} In fact  Eq.~\eqref{eq:normal_distribution} and  Eq.~\eqref{eq:elementary_exp} essentially are special case of the celebrated central limit theorem (CLT). The reader acquainted with the CLT is reminded, that the operations of multiplication and convolution in function space interchange their roles when subjected to a Fourier transformation. There is however no need to discuss the details of the proof of the CLT here: Luckily, our mechanical setting allows for a pedestrians approach to verify  Eq.~\eqref{eq:normal_distribution}. \\

The proof starts with the observation that the normalized bending profile of a fully concentrated load (only the right hand side of the symmetric profile is given),
\begin{equation}
    g(\xi) = (1-2\xi)^2 (1+4\xi) \quad , \quad 0 \le \xi \le \frac{1}{2}
\end{equation}
and the normalized bending profile of the fully distributed, i.e. constant load,

\begin{equation}
     h(\xi) = \left( 1 - 4 \xi^2 \right)^2
\end{equation}

provide an upper and a lower bound for $\phi_n(\xi)$,
\begin{equation}\label{eq:phi_bounds}
 G_n(\xi) \ge \phi_n(\xi) \ge H_n(\xi) \ .
\end{equation}
Here $G_n(\xi)$ and  $H_n(\xi)$  are defined analogously to Eq.~\eqref{eq:phi_n}, i.e. by replacing  $g(\xi)$ and $h(\xi)$ respectively for $\psi_0(\xi)$ in that equation (also the respective $\sigma$ needs to be calculated),
\begin{equation}  
    \begin{aligned}
        G_n(\xi) &= \frac{1}{\sqrt{2 \pi}}\left( 1-\frac{\xi}{\sqrt{6 n }}  \right)^{2 n}  \left( 1+\frac{2 \xi}{\sqrt{6 n }}  \right)^n , \\ 
        H_n(\xi) &= \frac{1}{\sqrt{2 \pi}}\left( 1 - \frac{\xi ^2}{4 n}\right)^{2 n} .
    \end{aligned}
\end{equation}
The relation 
Eq.~\eqref{eq:phi_bounds} 
is easily verified by establishing the assertion for 
$\mathit{n}=1$ 
first, and then using the positivity of the functions involved when raising to the 
$\mathit{n}$-th 
power. Note that the relation 
Eq.~\eqref{eq:phi_bounds} 
also is invariant under the scaling of the 
$\xi$-axis, 
required when progressing from 
$\mathit{n}$ to $\mathit{n}+1$. 
Computing the limiting function of the sequence
$\{\mathit{H}_\mathit{n}(\xi)\}_{n \in \mathbb{N}} $  
is a simple application of 
Eq.~\eqref{eq:elementary_exp},
\begin{equation}
    \begin{aligned}
         \lim_{n\to\infty} H_n(\xi) 
         &=\lim_{n\to\infty} \frac{1}{\sqrt{2 \pi}}\left( 1- \frac{\xi^2}{4 n}\right)^{2n} \\
         &=\lim_{m\to\infty} \frac{1}{\sqrt{2 \pi}}\left( 1- \frac{\xi^2}{2m}\right)^{m} \\
         &=\frac{1}{\sqrt{2\pi}} \exp \left( -\frac{\xi^2}{2}\right).
    \end{aligned}
\end{equation}
Computing the limiting function of the sequence 
$\{\mathit{G}_{\mathit{n}}(\xi)\}_{n \in \mathbb{N}} $  
is little more challenging,
\begin{equation}
    \begin{aligned}
               \lim_{n\to\infty} G_n(\xi) &=\lim_{n\to\infty} \frac{1}{\sqrt{2 \pi}}\left( 1- \frac{\xi^2}{2n} + \frac{\xi^3}{3 \sqrt{6 n^3} } \right)^n \\
             &=\lim_{n\to\infty} \frac{1}{\sqrt{2 \pi}} \left( 1- \frac{\xi^2}{2n} \right)^n \left( 1 +O\left(\frac{1}{n} \right)^\frac{1}{2}  \right)\\
              &=\frac{1}{\sqrt{2\pi}} \exp \left( -\frac{\xi^2}{2}\right).
             \end{aligned}
\end{equation}
Dini's theorem\cite{Forster.2017} asserts the uniformity of the convergence in both cases. Now since both, the upper and the lower bound of $\Phi_n(\xi)$ uniformly converge to the Gaussian, the same holds true for the sequence 
$\{\Phi_n(\xi)\}_{n \in  \mathbb{N}} $ 
itself, establishing 
Eq.~\eqref{eq:normal_distribution}. \\
%

Before ending this section, we would like to highlight that Chebyshev's general argument works in the domain of elasto-mechanics far beyond the simple case presented here and does not require any kind of symmetry. That is because Chebyshev essentially exploits the fact that Hermite polynomials form a complete base of the Hilbert space of functions over the reals, that are square integrable with respect to the measure defined by the Gauss bell curve.
\subsection{The Edgeworth expansion}
%
%
%
For the evaluation of the Coulomb integral $f_0(z)$ we need to know how exactly $\Phi_n(\xi)$ approaches Gauss' bell curve as n grows larger. The answer is provided by the famous Edgeworth expansion: Following the ideas of F.Y. Edgeworth,  Eq.~\eqref{eq:normal_distribution} warrants the existence of an asymptotic expansion of the form\cite{Edgeworth.1905, Wallace.1958}
\begin{equation}\label{eq:edgeworth_expansion}
   \Phi_n(\xi) = \frac{1}{\sqrt{2 \pi}} \exp\left( -\frac{\xi^2}{2}\right) \times 
   \left(
   1 
   - \frac{c_1(\xi)}{n}
   + \frac{c_2(\xi)}{n^2}
   +O\left(\frac{1}{n} \right)^3
   \right).
\end{equation}
The explicit version of this asymptotic expansion, is obtained by expanding $\Phi_{\mathit{n}}(\xi)$ in a Taylor series at $\mathit{n}=\infty$ in powers of $\mathit{n}^{-1}$. The computation of the respective Taylor coefficients is enabled by the use of  Eq.~\eqref{eq:normal_distribution} and of 
\begin{equation}\label{eq:Edgeworth_1st_power}
    \sqrt{2\pi} \, \Phi_1(\xi)=
    1
    -\frac{\xi^2}{2}
    + \frac{\mu_4 \, \xi^4}{24}
    - \frac{\mu_6 \, \xi^6}{720}
    + O(\xi)^8 \, .
    \\
\end{equation} 
Here we have introduced the following abbreviations related to the derivatives of order 2k,
\begin{equation}\label{eq:even_moments}
    \mu_{2k} = 
    \frac{(-1)^k}{\sigma^{2k}} \frac{\psi_0^{(2k)}(0)}{\psi_0 (0)} \, ,
\end{equation}
\begin{equation}\label{eq:moments}
\mu_{4k} 
= \mu_{4k+2} 
= \left(
        \frac
            {\cosh(\beta_0 / 2)- \cos(\beta_0 / 2)}
            {\cosh(\beta_0 / 2)+ \cos(\beta_0 / 2)} 
    \right)^{2k}\ .
\end{equation}
The sequence of integers appearing in Eq.~\eqref{eq:Edgeworth_1st_power} is the sequence of the non prime factorials; this jointly with Eq.~\eqref{eq:moments} implies that the expansion Eq.~\eqref{eq:Edgeworth_1st_power} is absolutely convergent within an infinite radius of convergence. Merten's theorem regarding Cauchy products\cite{Konigsberger.2001} therefore ensures that any integer power of Eq.~\eqref{eq:Edgeworth_1st_power}, required for the evaluation of Eq.~\eqref{eq:phi_n} exists, also possessing an infinite radius of convergence. 

%

The first two coefficients of the Edgeworth expansion obtained following the route outlined here are,

\begin{equation}\label{eq:edgeworth_coefficients}
   \begin{aligned}
    c_1(\xi)&=\frac{-3 +\mu_4}{24} \, \xi^4 \, ,\\
    c_2(\xi)&=\frac{(-3 +\mu_4)^2}{1152} \, \xi^8 \,
             +\frac{-30 + 15\mu_4 - \mu_6}{720} \, \xi^6 .
\end{aligned}
\end{equation}

Finally we would like to add that in the setting of analytical probability theory, the 
$\Phi_1(\xi)$ 
plays the role of the characteristic function of a probability density and the
$\{\mu_\mathit{n}\}_{n \in \mathbb{N}}$
are its respective moments. In our case, the Fourier transform of 
$\Phi_1(\xi)$ ,
which should be a probability density, can adopt negative values. It only is asymptotically a non negative function. So the notions of analytical probability theory, strictly speaking, do not apply. However the line of arguments of Chebyshev and Edgeworth still hold under our somewhat weaker conditions, as we have explicitly shown above.

\subsection{Evaluating the Coulomb integral}

\begin{figure*}[hbtp]
	\centering
	(a)\includegraphics[height = 6.5 cm]{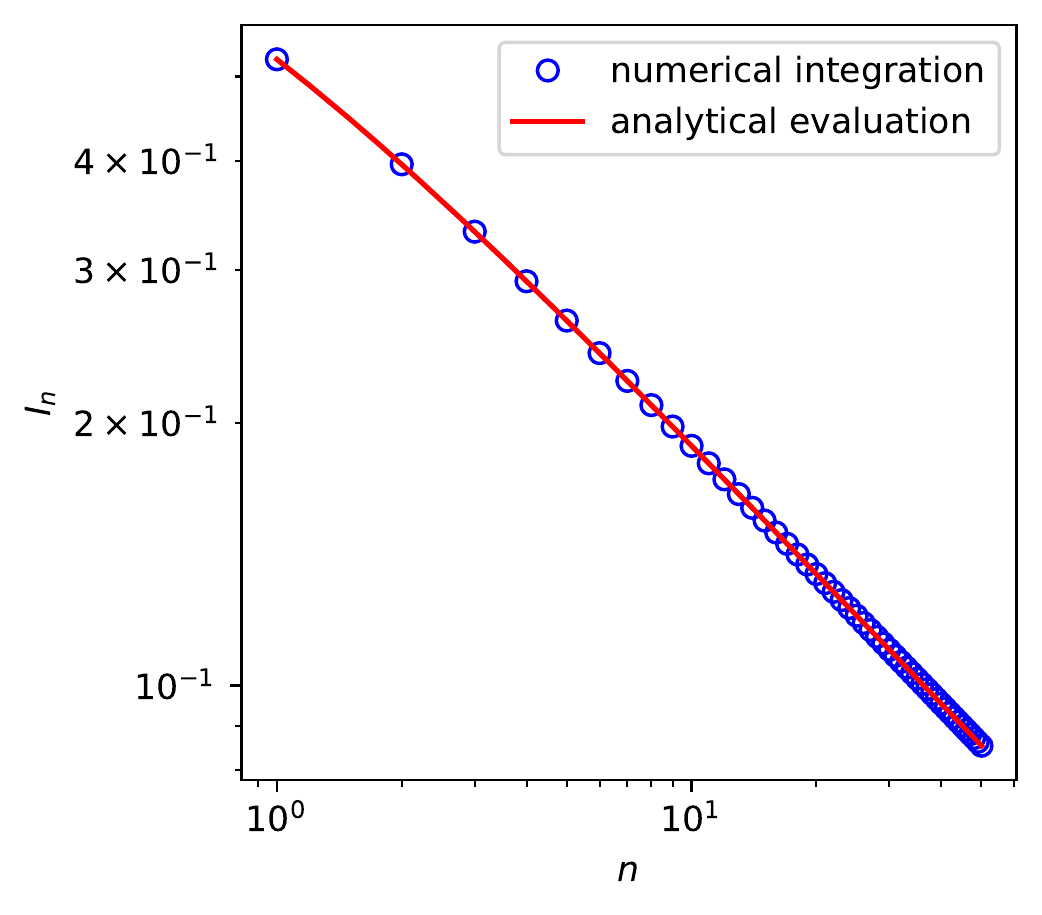}
	(b)\includegraphics[height = 6.5 cm]{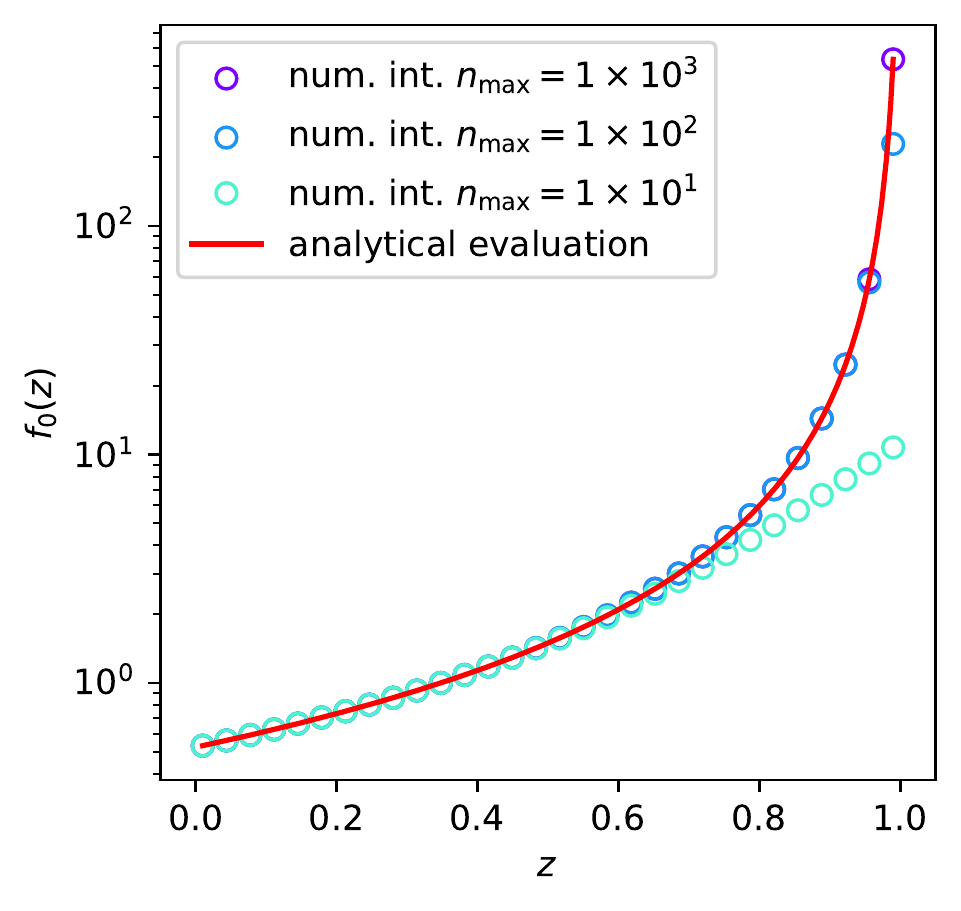}
	\caption{
	(a) Comparison of two methods to compute the integrals $I_n$ according to Eq.~\eqref{eq:In_integral_1}. Open circles mark the results from direct numerical integration. The red line is the result of the Chebyshev-Edgeworth formula Eq.~\eqref{eq:Cby_In}.
	(b) Comparison of two methods to compute the Coulomb integral $f_0(z)$ as defined by Eq.~\eqref{eq:Coulomb_f0_def}. Open circles mark the results from evaluating the sum Eq.~\eqref{eq:f0_In_series} up to a certain maximum number of terms $n_{max}$ by direct numerical integration. The red line is the Chebychev-Edgeworth projection Eq.~\eqref{eq:f0_Li} of the Coulomb force, neglecting the remainder $\Delta_{f_0}(z)$ .
	}
	\label{fig:Chebyshev_Edgeworth}
\end{figure*}

In this section we evaluate  
Eq.~\eqref{eq:In_integral_2} 
and perform the summation 
Eq.~\eqref{eq:f0_In_series}. 
The last subtlety to cope with, are the finite boundaries of the integral 
Eq.~\eqref{eq:In_integral_2}.
While it is perfectly possible to analytically perform the integration within these finite boundaries and expand the results in terms of 
$\mathit{n}^{-1}$, 
little is gained by this tedious exercise. Truncating the integrand at order 
$\mathit{O}\left(\mathit{n} \right)^{-3}$ 
according to 
Eq.~\eqref{eq:edgeworth_expansion} 
and extending the integration boundaries of  
Eq.~\eqref{eq:In_integral_2} to infinity generates an overall error, which is negligible for all practical purposes, as we will show in Eq.~\eqref{eq:In_error} below. Therefore we evaluate 
Eq.~\eqref{eq:In_integral_2} 
in the form,
\begin{equation}\label{eq:In_integral_3}
  I_n = \frac{\sqrt{2 \pi}}{\sigma \sqrt{n}} 
  \int_{-\infty}^{+\infty} \Phi_n(\xi) \, \mathrm{d}\xi \,
  \pm \bar{\Delta}_n \ .
\end{equation}
Inserting 
Eq.~\eqref{eq:edgeworth_expansion} 
and  
Eq.~\eqref{eq:edgeworth_coefficients}  
into 
Eq.~\eqref{eq:In_integral_3} yields 
\begin{equation}\label{eq:Cby_In}
    I_n = \frac{\sqrt{2 \pi}}{\sigma} \times \\
    \left( 
    \frac{1}{n^\frac{1}{2}}
    + \frac{\mu_4 - 3}{8 \, n^{\frac{3}{2}}} 
    + \frac{75 - 90\mu_4 + 35 \mu_4^2 - 8\mu_6 }{384 \, n^{\frac{5}{2}}} \right)\\
   + \Delta_n ,
\end{equation}
where the remainder
$\Delta_{\mathit{n}}$
in Eq.~\eqref{eq:Cby_In}
accounts for both simplifications mentioned above. An upper bound for  $\Delta_{\mathit{n}}$ can easily be found upon noticing that the maximum remainder occurs for $\mathit{n}=2$. 
Taylor's remainder theorem then asserts that according to Eq.~\eqref{eq:edgeworth_expansion}  
and 
Eq.~\eqref{eq:In_integral_3} 
the remainder decays at least with the power 
$\mathit{n}^{\frac{5}{2}}$,

\begin{equation}\label{eq:In_error}
     0 < \Delta_n
     \leq \Delta_2 \left( \frac{2}{n} \right)^{\frac{5}{2}}
     <  \frac{1}{1956 \; n^{\frac{5}{2}}}\, .
\end{equation}
The excellent accuracy of the expansion 
Eq.~\eqref{eq:Cby_In} 
for 
$I_{\mathit{n}}$ 
is apparent from 
Fig.~\ref{fig:Chebyshev_Edgeworth}a.

To compute the Coulomb integral $f_0(z)$, we need, last not least, to perform the summation according to Eq.~\eqref{eq:f0_In_series}. The result is given in Eq.~\eqref{eq:f0_Li}, which we will call the Chebyshev-Edgeworth projection of the Coulomb force,
\begin{widetext}
\begin{equation}
      f_0(z) = \frac{\sqrt{2 \pi}}{\sigma} \frac{1}{z} \,
    \left( 
    \mathrm{Li}_{-\frac{1}{2}}(z)
    + \frac{\mu_4 - 3}{8} \mathrm{Li}_{\frac{1}{2}}(z)
    + \frac{75 - 90\mu_4 + 35 \mu_4^2 - 8\mu_6 }{384} \mathrm{Li}_{\frac{3}{2}}(z)
    \right)
    + \Delta_{f_0}(z)  \, .
    \label{eq:f0_Li}
\end{equation}
\end{widetext}

In  Eq.~\eqref{eq:f0_Li} the function $\mathrm{Li}_s(z)$ denotes Jonquière's  poly-logarithm\cite{Jonquiere.1889}, defined for all $\left| z \right| < 1$ and for all $s\in \mathbb{C}$ as
\begin{equation}\label{eq:Li_definition}
    \mathrm{Li}_s(z) = \sum_{n=1}^{\infty} \frac{z^n}{n^s} \, .
\end{equation}
Note the relation with Riemann's zeta function \cite{Riemann.1859} $\zeta(z)$ relevant to us,
\begin{equation}
     \mathrm{Li}_s (1) = \zeta(s)\, .
\end{equation}
Based on Eq.~\eqref{eq:In_error} and on Eq.~\eqref{eq:Li_definition} we can evaluate an upper bound for the remainder $\Delta_{f_0}(z)$ 
\begin{equation}\label{eq:f0_remainder}
    \begin{aligned}
        0 < \Delta_{f_0}(z) 
        &< \frac{1}{1956 \, z} \mathrm{Li}_{\frac{3}{2}}(z) \, \\ 
        \frac{1}{1956 \, z}    \mathrm{Li}_{\frac{3}{2}}(z)
        &\leq \frac{1}{1956}   \mathrm{Li}_{\frac{3}{2}}(1)
        < \frac{1}{1236} . \\
    \end{aligned}
\end{equation}

Fig.~\ref{fig:Chebyshev_Edgeworth}b shows that the Chebyshev-Edgeworth projection produces excellent results for the Coulomb integral  Eq.~\eqref{eq:Coulomb_f0_def}. The underlying reason is that the  Chebyshev-Edgeworth expansion maintains the exact singularity structure of the Coulomb integral. This does not hold true for the approach of Younis et al. \cite{M.I.Younis.2003b}. 

\subsection{The singular structure of the Coulomb Integral}

The Chebyshev-Edgeworth projection formula Eq.~\eqref{eq:f0_Li} may actually appear a bit awkward, from a practitioners point of view. In the sequel we will seek to improve this. The key information contained in Eq.~\eqref{eq:f0_Li} is how exactly to deal with the Coulomb force: The zero-mode approximation is a projection of the beam equation Eq.~\eqref{eq:full_beam1} from the infinite-dimensional Hilbert space of all Euler-Bernoulli eigenmodes onto the one-dimensional subspace, spanned by the zero-mode $\psi_0(z)$ only. A priori, it is far from obvious what this means for the Coulomb force. Note that this kind of question arises in any type of Galerkin procedure applied to Eq.~\eqref{eq:full_beam1}. Eq.~\eqref{eq:f0_Li} allows us to give the answer in case of the zero-mode approximation, leading to a more practical version of our projection formula.

The contact singularity of the Coulomb force obviously causes a singularity of the Coulomb integral $f_0(z)$ at $\mathit{z}=1$. The information about this singularity  is entirely contained in the poly-logarithms $\mathrm{Li}_{-\frac{1}{2}}(z)$ and $\mathrm{Li}_{+\frac{1}{2}}(z)$; all poly-logarithms of larger index are regular,
\begin{equation}\label{eq:Li_asymptotics}
    \begin{aligned}
        \mathrm{Li}_{-\frac{1}{2}}(z) \,\,
                        &= 
                             \mathrm{Li}_{-\frac{1}{2}}(1)
                         +  \frac{ \sqrt{\pi}}{ 2 (1-z)^{\frac{3}{2}}} 
                         -  \frac{\sqrt{3\pi}}{8 (1-z)^{\frac{1}{2}}} 
                         +  O(1-z)^{\frac{1}{2}} \ , \\
     \mathrm{Li}_{\frac{1}{2}}(z)\quad \,
                        & = 
                           \mathrm{Li}_{\frac{1}{2}}(1)
                         + \frac{\sqrt{\pi}}{(1-z)^{\frac{1}{2}}} 
                         + O(1-z)^{\frac{1}{2}} \ , \\
     \mathrm{Li}_{n + \frac{1}{2}}(z) 
                        &= 
                          \mathrm{Li}_{n + \frac{1}{2}}(1)
                         + O(1-z)^{\frac{1}{2}} \quad ,\quad 0 < n \in \mathbb{N} \,.
    \end{aligned}
\end{equation}
%
%
%

This local analysis reveals that $f_0(z)$ as defined in Eq.~\eqref{eq:Coulomb_f0_def} has the singular structure,
\begin{equation}\label{eq:f0_singularity}
   \begin{aligned}
        f_0(z)   &= \hat{f}_0(z) + O(1-z)^{\frac{1}{2}} \ , \\  
    \hat{f}_0(z) &= \mathit{a} 
                   + \frac{\mathit{b}}{(1-z)^{\frac{1}{2}}} 
                   + \frac{\mathit{c}}{(1-z)^{\frac{3}{2}}}\ . 
    \end{aligned}
\end{equation}
We now wish to find an algebraic approximation of the form $\hat{f}_0(z)$ to Eq.~\eqref{eq:f0_Li} that is as accurate as possible over the  entire range $0 \leq z \leq 1$.  This means we give up a little bit of the achieved accuracy at the singularity, in exchange for a global approximation, that pointwise has a relative error small enough for all practical purposes. To this end we demand that $\mathit{a}$, $\mathit{b}$ and $\mathit{c}$  minimize the distance between  $f_0(z)$ and its algebraic approximation $\hat{f}_0(z)$ with respect to a suitable norm in function space. The challenge here is the isolated singularity at $\mathit{z} = 1$. The associated lack of integrability can however be mended by introducing an apt non negative weight function $\mathit{r}(z)$. The weight function should be selected such that it has a zero of sufficiently high degree compensating the singularity. Having said this, we choose the coefficients $\mathit{a}$, $\mathit{b}$ and $\mathit{c}$ to minimize the functional
\begin{equation}\label{eq:S_functional}
    \mathcal {S}_r (\mathit{a}, \mathit{b}, \mathit{c}) =
    \int_0^1 \mathit{r}(z) \left (f_0(z) - \hat{f}_0(z) \right)^2 dz \, .
\end{equation}
A suitable $\mathit{r}(z)$ ensures the existence of this functional and of its Hessian as a positive definite matrix. Conceptually, an optimal weight function simultaneously minimizes the relative error. In practice, our simplistic choice, justified in arrears  by Eq.~\eqref{eq:f0_hat_error}, is
\begin{equation}
    \mathit{r}(z) = (1-z)^3 .
\end{equation}
With this weight function the Hessian is 
\begin{equation}
    \mathcal{H}\mathit{e}\mathit{s}\mathit{s} (\mathcal{S}_r) = 
    \left (
            \begin {array} {ccc} 
                       1/4 & 2/7 & 2/5 \\ 
                       2/7 & 1/3 & 1/2 \\ 
                       2/5 & 1/2 & 1 
            \\\end{array} 
    \right) \, .
\end{equation}
Accordingly, there is a uniquely defined minimum which is found solving for 
\begin{equation}\label{eq:algebr_linear_system}
\begin{aligned}
    \frac{\partial}{\partial\mathit{a}}\mathcal{S}_r(\mathit{a},\mathit{b},\mathit{c}) = 0 \, 
    &\Leftrightarrow \,
    H_a = \frac{\mathit{a}}{4} + \frac{2\mathit{b}}{7} + \frac{2\mathit{c}}{5} \\
    \frac{\partial}{\partial\mathit{b}}\mathcal{S}_r(\mathit{a},\mathit{b},\mathit{c}) = 0 \, 
    &\Leftrightarrow \,
    H_b= \frac{\mathit{2 a}}{7} + \frac{\mathit{b}}{3} + \frac{\mathit{c}}{2} \\
    \frac{\partial}{\partial\mathit{c}}\mathcal{S}_r(\mathit{a},\mathit{b},\mathit{c}) = 0 \, 
    &\Leftrightarrow \,
    H_c= \frac{\mathit{2 a}}{5} + \frac{\mathit{b}}{2} + \mathit{c}
    \end{aligned}
\end{equation}
The constants $H_a$, $H_b$ and $H_c$ are the integrals
\begin{equation}
    \begin{aligned}
          H_a &= \int_0^1 (1-z)^3 f_0(z) dz  \, ,\\
          H_b &= \int_0^1 (1-z)^{\frac{5}{2}} f_0(z) dz  \, ,\\
          H_c &= \int_0^1 (1-z)^{\frac{3}{2}} f_0(z) dz \, .\\
    \end{aligned}
\end{equation}
Using suitable integer fractions we find the targeted algebraic expansion for $\hat{f}_0(z)$ to be,
\begin{equation}\label{eq:f0_hat}
    \hat{f}_0(z) = \frac{1}{77} 
            - \frac{1}{38 (1-z)^{\frac{1}{2}}} 
            + \frac{15}{28(1-z)^{\frac{3}{2}}}  \, .
\end{equation}
The upper bound for the maximum relative error regarding this greatly simplified version of the Coulomb integral is easily computed analytically to be 
\begin{equation}\label{eq:f0_hat_error}
    \begin{aligned}
        \max_{0\leq z\leq 1} \left( 1 - \frac{\hat{f}_0(z)}{f_0(z)} \right) 
        &\leq  \lim_{z\to 1} \left( 1 - \frac{\hat{f}_0(z)}{f_0(z)} \right) \, , \\
        &= 1-\frac{15\sigma}{14 \sqrt{2}\pi} 
        <  \frac{1}{494} \, .
    \end{aligned}
\end{equation}
This is excellent for all practical purposes.
In summary we have shown that the Coulomb singularity of 
Eq.~\eqref{eq:full_beam1} 
transforms into the quite different singularity given by the asymptotic expansion of Eq.~\eqref{eq:f0_Li}, 
or for all practical purposes, by the global approximation 
Eq.~\eqref{eq:f0_hat}, 
when projected onto the one dimensional Hilbert subspace spanned by the Euler-Bernoulli zero-mode. To the best of our knowledge this is a completely new result of substantial practical relevance.

\subsection{Synopsis of the zero-mode LPM}\label{sec:synopsis}

The zero-mode approximation 
\begin{equation*}
    w(\xi,\tau) \approx z(\tau) \ \frac{\psi_{0}(\xi)}{\psi_{0}(0)} \, ,
\end{equation*}
developed in the previous section, leads upon careful treatment of the Coulomb singularity to the lumped parameter model, 
\begin{equation*}
    \frac{\partial^2}{\partial \tau^2} z 
    + c \frac{\partial}{\partial \tau} z  
    + k_0 \ z
    + \kappa \  z^3 
    = u^2 \hat{f}_0(z) \, , \\
\end{equation*}
\begin{equation}\label{eq:LPM}
    \hat{f}_0(z) = \frac{1}{77} 
            - \frac{1}{38 (1-z)^{\frac{1}{2}}} 
            + \frac{15}{28(1-z)^{\frac{3}{2}}} 
            \, ,
\end{equation}
\begin{equation*}
    \kappa = \alpha_1 \left( \frac{\chi_0}{\psi_0(0)} \right)^2 , \
       k_0 =\lambda_0 + N \chi_0 , \
\end{equation*}
\begin{equation*}
          u = \psi_0(0) \sqrt{\alpha_2} \, v \, ,
\end{equation*}
\begin{equation*}
  \alpha_1 = 6 \left( \frac{g}{t} \right) ^2 \, , \,
  \alpha_2 = \frac{6 \epsilon l^4}{E t^3 g^3} \, . 
\end{equation*}
Here $l$, $t$, and $E$ denote length, thickness and  Young's modulus of the beam. $g$ is the electrode gap. The definitions of $\psi_0(0)$ and $\chi_0$ can be found in Eq.~\eqref{eq:zero_mode_ansatz}, Eq.~\eqref{eq:beta_definition} and Eq.~\eqref{eq:chi0}.
For higher precision, Eq.~\eqref{eq:f0_Li} or any refinement thereof, can be used instead of Eq.~\eqref{eq:f0_hat}.
The bifurcation diagram, showing the static deflection of the beam center as a function of the drive voltage, is obtained as the set of all points in the $(u,z)$ plane, solving the purely algebraic equation
\begin{equation}
    \kappa \, z^3 + k_0 \, z = u^2 \hat{f}_0(z) \, .
    \label{eq:bifurcation_diagram}
\end{equation}
Eq.~\eqref{eq:bifurcation_diagram} is best used by looking upon the voltage $u$ as a function of the deflection, i.e. $u=u(z)$.  The static pull-in deflection $z_{PI}$ is reached at the critical point where 
\begin{equation}
   \frac{\partial u}{\partial z}(z_{PI}) = 0 \, .
   \label{eq:pull_in_condition}
\end{equation}
This condition is conveniently exploited by taking the inverse of the logarithmic derivative of  Eq.~\eqref{eq:bifurcation_diagram}. As shown in   Eq.~\eqref{eq:DLogf0Inv_error_bound} below, within very small error margins, the inverse of the logarithmic derivative of the Coulomb integral is a linear function of the deflection amplitude z, 
\begin{equation}\label{eq:DLogf0Inv}
  \left( \frac{\partial}{\partial z} \log (f_0(z))\right) ^{-1}
   = \frac{64}{97} - \frac{44}{67}z  - \Delta_{LOG}(z) \, . \\ 
\end{equation}
The approximation Eq.~\eqref{eq:DLogf0Inv} is obtained upon inserting Eq.~\eqref{eq:f0_In_series} into the left hand side of Eq.~\eqref{eq:DLogf0Inv} and performing a Taylor expansion. The maximum error occurs at $z=1$ where the left hand side of Eq.~\eqref{eq:DLogf0Inv} vanishes, due to the nature of its singularity as exhibited in  Eq.~\eqref{eq:f0_singularity}. This puts a tight absolute bound on the remainder $\Delta_{LOG}(z)$, 
\begin{equation}
    0 \, \le \,  \Delta_{LOG}(z) < \frac{1}{325}\, .
    \label{eq:DLogf0Inv_error_bound}
 \end{equation}

 It should be emphasised, that the derivation of  Eq.~\eqref{eq:DLogf0Inv} does not require using   Eq.~\eqref{eq:f0_hat} or any other approximation discussed in this paper. The absolute upper bound of the remainder $\Delta_{LOG}(z)$ is therefore not affected by any choice or error estimate made elsewhere.
The highly effective approximation Eq.~\eqref{eq:DLogf0Inv} leads to a simple algebraic equation for the practical evaluation of the pull-in deflection $z_{PI}$,
\begin{equation}\label{eq:z_Pull_In}
   \frac{\kappa \, z_{PI}^3 + k_0 \, z_{PI}}{3 \kappa \, z_{PI}^2 + k_0} 
   \approx \frac{64}{97} - \frac{44}{67}z_{PI}  \, .
\end{equation}
A first easy conclusion that can be drawn from Eq.~\eqref{eq:z_Pull_In} is that the pull-in deflection of a Coulomb actuated clamped-clamped Euler Bernoulli beam varies within the limits
\begin{equation}\label{eq:z_PI_bounds}
        0.3982 \leq z_{PI} \leq 0.6664 \, .
\end{equation}
\textcolor{black}{
The lower bound of Eq.~\eqref{eq:z_PI_bounds}  is obtained as the limiting case of Eq.~\eqref{eq:z_Pull_In}, where the stress stiffening (Duffing) coefficient $\kappa$ vanishes. Likewise, the upper bound of Eq.~\eqref{eq:z_PI_bounds}  results from Eq.~\eqref{eq:z_Pull_In} in case of an infinitely large $\kappa$.}
Within the realm of Euler-Bernoulli theory, these boundaries are independent of the shape of the beam cross section.
\textcolor{black}{
While this fact certainly is known from numerical studies\cite{Melnikov.2021}, it is derived here based on an analytical model, probably for the first time.}

\begin{figure}[hbtp]
	\centering
	\includegraphics[height = 6.5 cm]{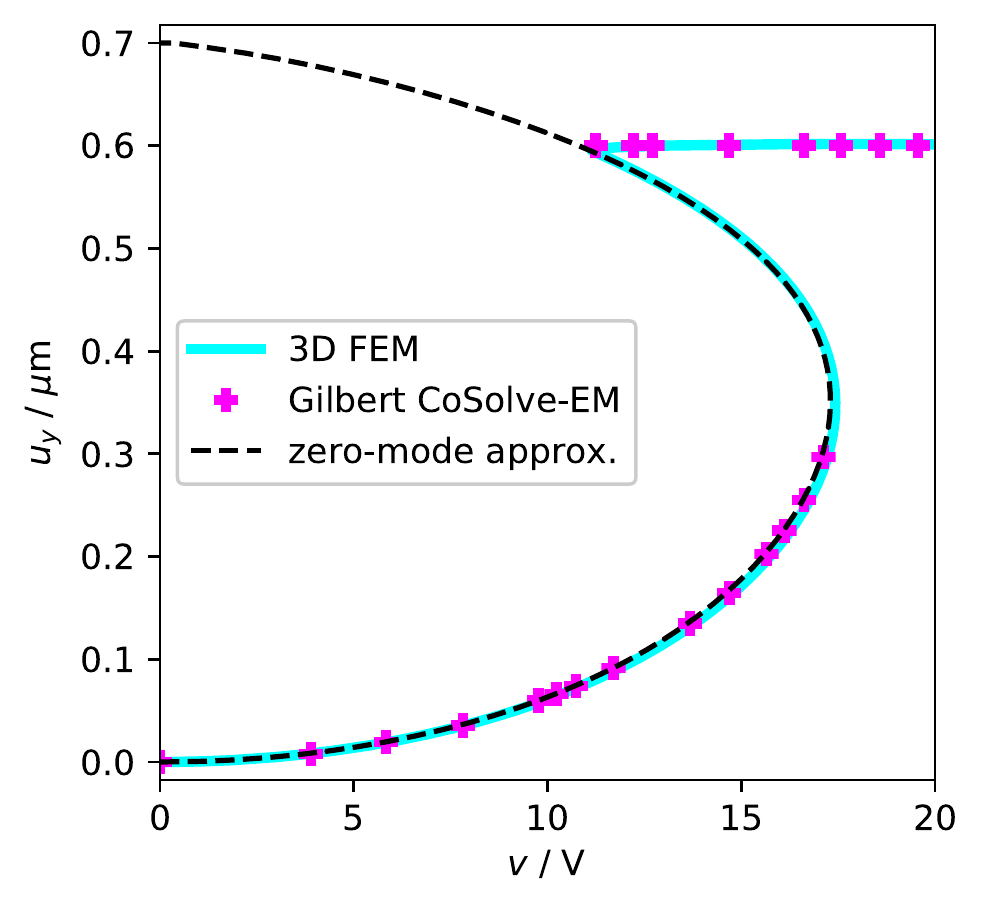}
	\caption{
	Comparison of the deflection curve obtained from the zero-mode approximation  Eq.~\eqref{eq:LPM} to the 3D simulation results from literature (Gilbert CoSolve-EM \cite{Gilbert.1996} and 3D FEM \cite{Melnikov.2021}).
	}
	\label{fig:Gilbert}
\end{figure}

\begin{figure*}[hbtp]
	\centering
	(a)\includegraphics[height = 6.5 cm]{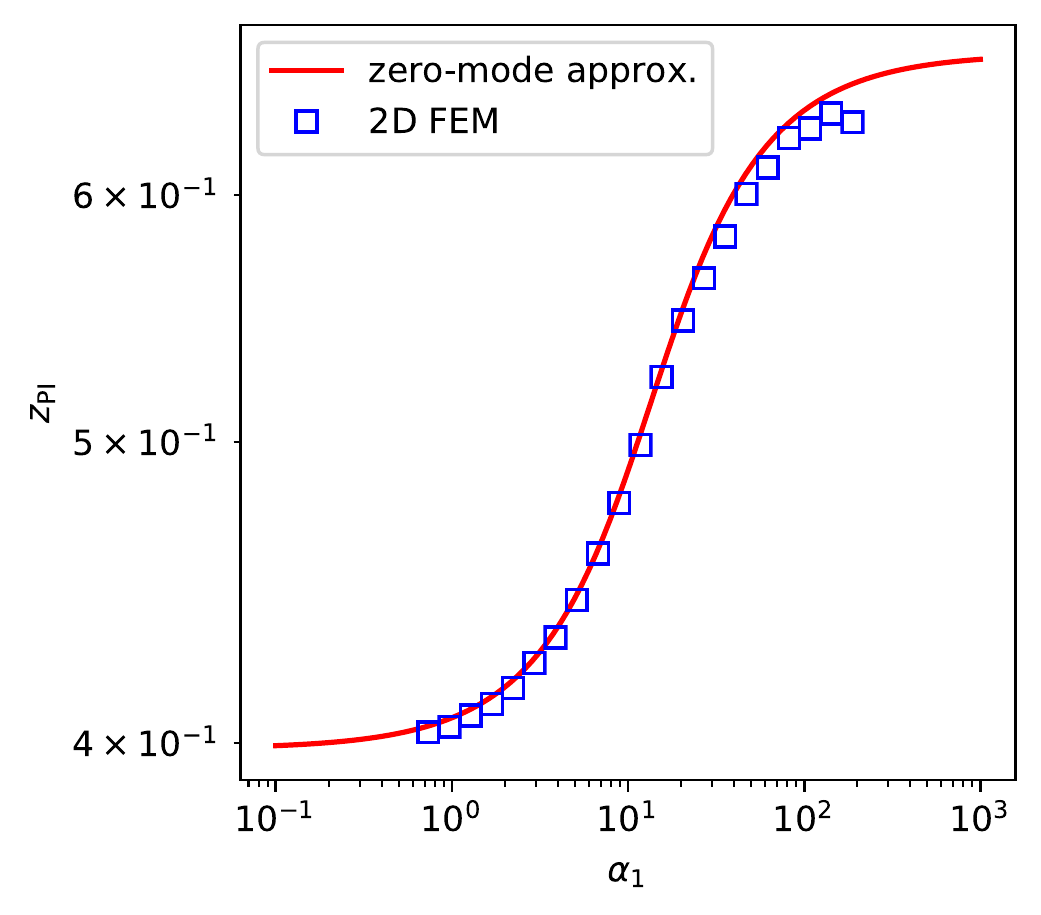}
	(b)\includegraphics[height = 6.5 cm]{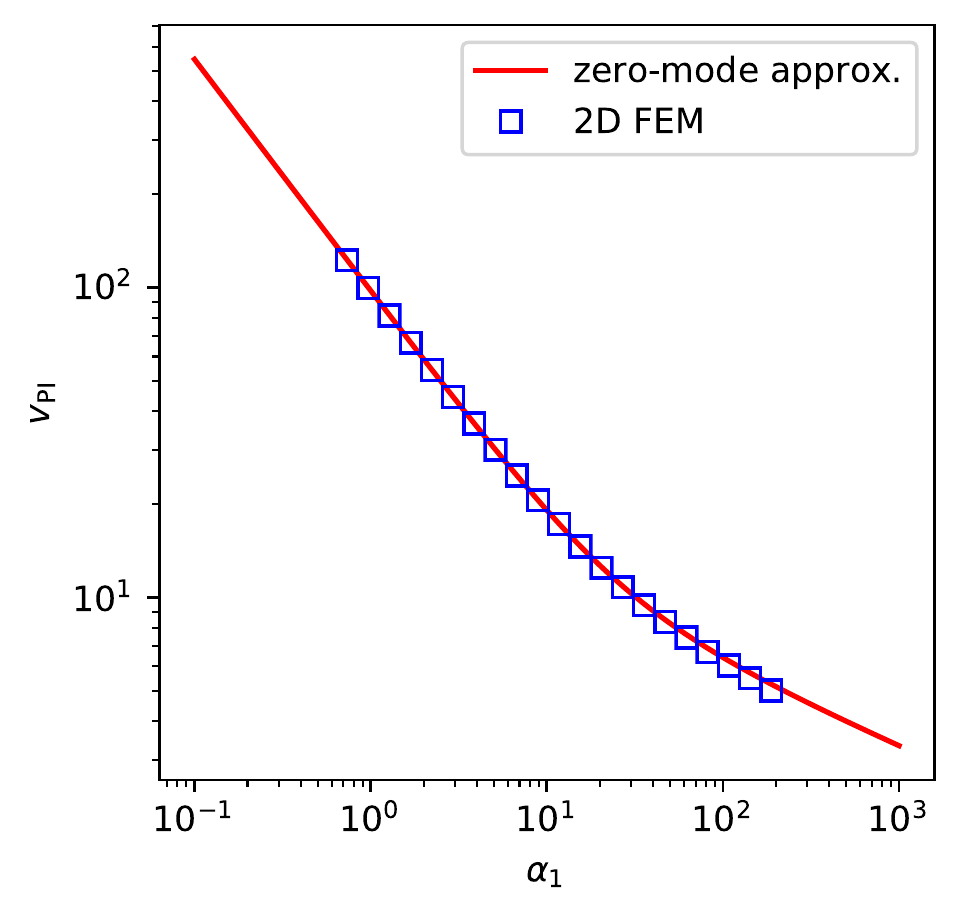}
	\caption{
	(a) The pull-in deflection z of a Coulomb actuated beam obtained by ANSYS (blue squares) and by  zero-mode approximation Eq.~\eqref{eq:z_Pull_In} as a function of $\alpha_1$ (solid red line).
	(b)The pull-in voltage of a Coulomb actuated beam simulated by ANSYS (blue squares) in comparison to simulations based on the zero-mode approximation Eq.~\eqref{eq:z_Pull_In} and  Eq.~\eqref{eq:bifurcation_diagram}  as a function of $\alpha_1$ (solid red line).
	}
	\label{fig:pull-in data}
\end{figure*}

Once we know $z_{PI}$, we can find the respective pull-in voltage $u_{PI}$ using Eq.~\eqref{eq:bifurcation_diagram}. The simple recipe  presented in this section, requires little more than a spreadsheet or a pocket calculator to compute the pull-in data and the entire bifurcation diagram, with the astonishing numerical accuracy exhibited in Fig.~\ref{fig:Gilbert}, Fig.~\ref{fig:pull-in data} and Fig.~\ref{fig:onions}.
%
\subsection{LPM analysis of the beam used by Gilbert \textit{et al.}}\label{sec:gilbert}
As a first application of our single degree of freedom LPM, we use the zero-mode approximation to compute the equilibria of the Coulomb actuated prismatic Euler-Bernoulli beam studied by Gilbert \textit{et al.} \cite{Gilbert.1996}. For this exercise we apply  the formulae compiled in Section~\ref{sec:synopsis}. Gilbert used the geometrical dimensions:
beam length $l = \unit[80]{\mu m}$, 
beam width $w = \unit[10]{\mu m}$, 
beam thickness $t = \unit[0.5]{\mu m}$, 
electrostatic gap $g = \unit[0.7]{\mu m}$, and
stop layer $s = \unit[0.1]{\mu m}$.
For silicon, Gilbert used an isotropic stiffness with a Young's modulus of $E = \unit[169]{GPa}$ and a Poisson ratio of $\nu = 0.25$.

The zero-mode results are compared to the results of Gilbert \textit{et al.} \cite{Gilbert.1996} and to the 3D ANSYS simulation by Melnikov \textit{et al.} \cite{Melnikov.2021} in Fig.~\ref{fig:Gilbert}.
Obviously there is a very good agreement between our zero-mode approximation based on the Chebyshev-Edgeworth expansion and the results of Gilbert \textit{et al.}\cite{Gilbert.1996} and Melnikov \textit{et al.}\cite{Melnikov.2021} The deflection profile, the pull-in voltage, and the pull-out voltage can be reliably determined using our method.

\subsection{Comparison with the numerical and experimental results of Melnikov \textit{et al.}}\label{sec:melnikov}

Melnikov \textit{et al.} \cite{Melnikov.2021} used a continuation method to extend the reach of FEM simulations to the entire bifurcation diagram of  Coulomb actuated  prismatic clamped-clamped Euler-Bernoulli beams, including all stable and unstable equilibria. They calculated the respective bifurcation diagrams and pull-in voltages for micro-beams  with a length of  $l = \unit[80]{\mu m}$, a thickness range between  $t = \unit[0.12]{\mu m}$ and $t = \unit[2]{\mu m}$ and an electrode gap of $g = \unit[0.7]{\mu m}$. 

\begin{figure*}[hbtp]
	\centering
	(a)\includegraphics[height = 5.5 cm]{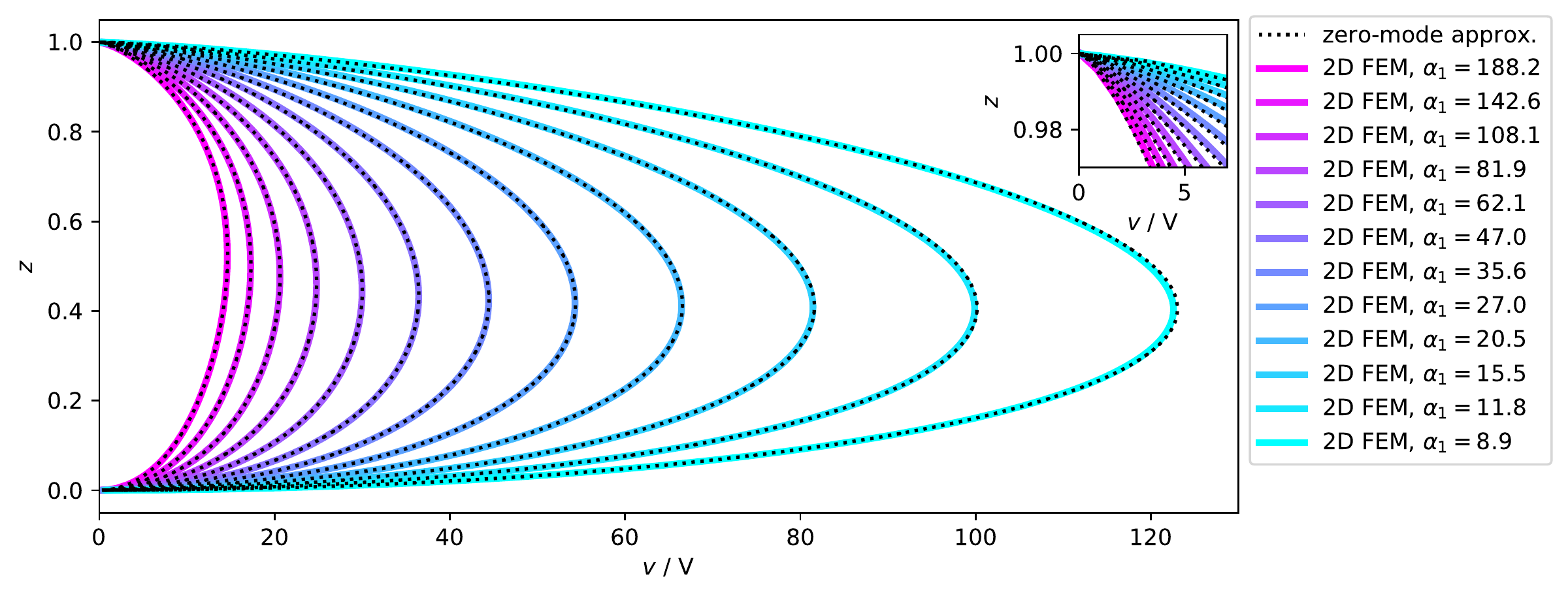}\\
	(b)\includegraphics[height = 5.5 cm]{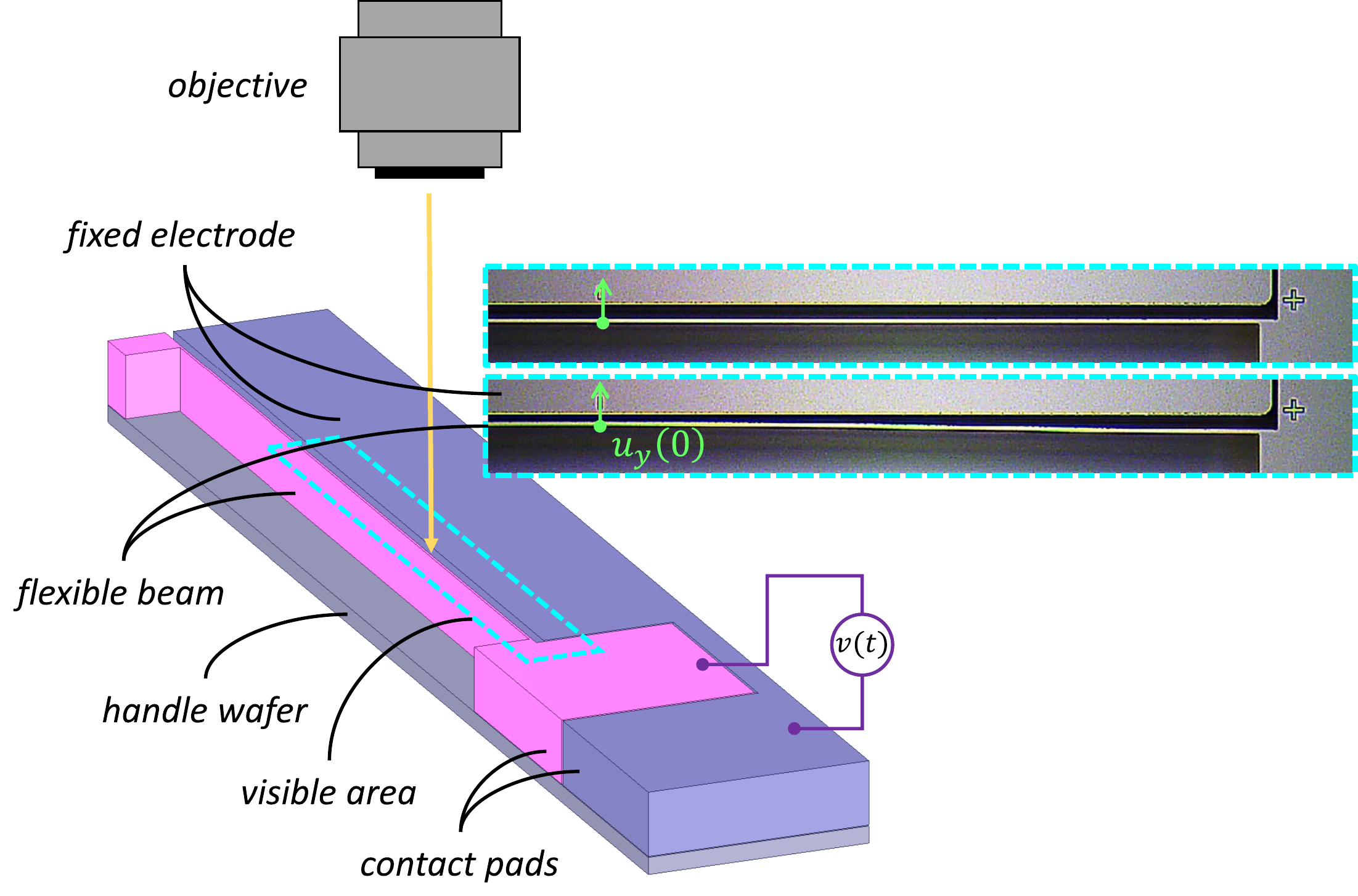}
	(c)\includegraphics[height = 5.5 cm]{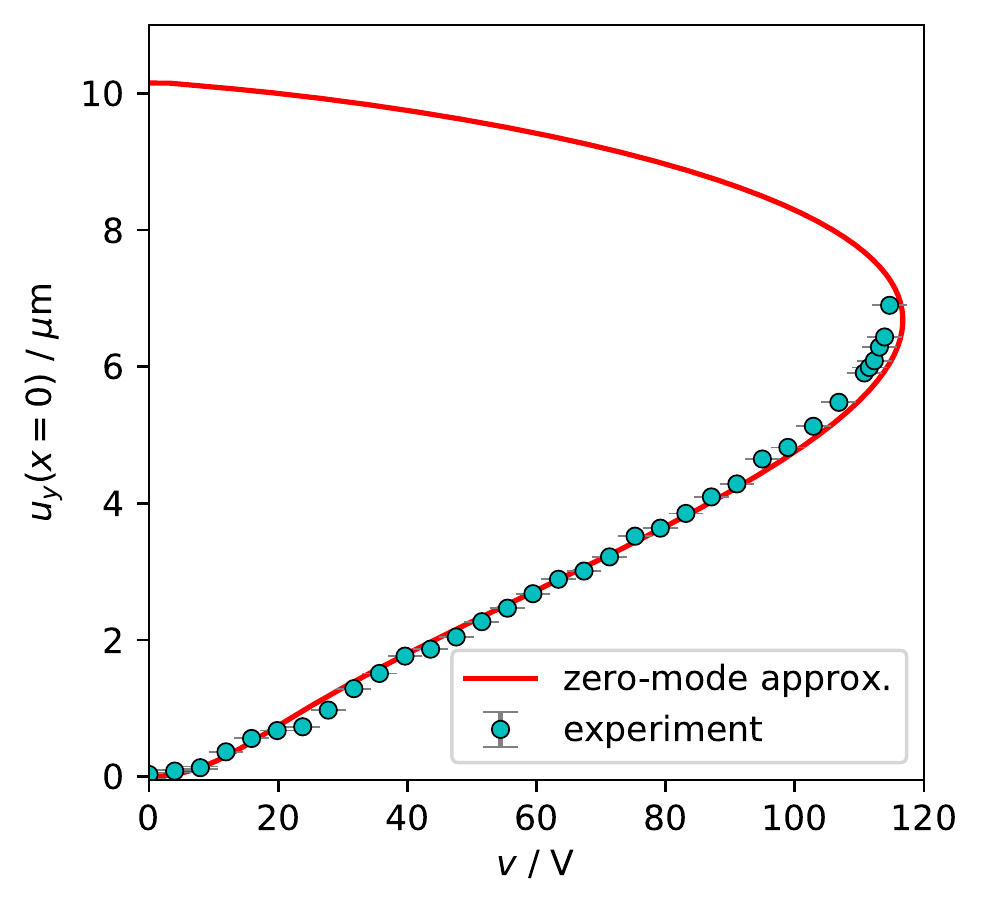}
	\caption{
	(a) The equilibria of a Coulomb actuated beam as obtained by ANSYS (colored solid lines) and by  the zero-mode approximation Eq.~\eqref{eq:LPM} (dotted black lines) for various thicknesses $t$.
	(b) The experimental setup and example frames used for the deflection measurement.
	(c) Measured equilibria of a Coulomb actuated beam (filled green circles) compared to the zero-mode approximation (solid red line) according to Eq.~\eqref{eq:LPM} as a function of the voltage.}
	\label{fig:onions}
\end{figure*}

Fig.~\ref{fig:pull-in data} shows the pull-in deflection and the pull-in voltage, respectively. These graphs demonstrate the excellent match of the zero-mode approximation and the FEM results.
Additionally, Fig.~\ref{fig:onions} reveals an almost perfect agreement between FEM results and the zero-mode approximation, regarding the entire deflection profiles, including their unstable branches. We note that the solution close to the contact singularity at $z=1$ is correctly reproduced using a single mode.

Melnikov \textit{et al.} \cite{Melnikov.2021} scrutinized their findings by runing a \gls{MEMS} experiment. The basic experimental set-up is shown in Fig.~\ref{fig:onions}b. A clamped-clamped \gls{MEMS} micro-beam of length $l = \unit[1000]{\mu m}$,  width $w = \unit[75]{\mu m}$ and with a measured thickness of $t = \unit[2.47]{\mu m}$ was manufactured on a Bonded Silicon on Insulator (BSOI) wafer, to perform in-plane movements. The beam is Coulomb actuated by a planar electrode positioned in front of the beam at a distance of $g = \unit[10.15]{\mu m}$ (fitted electrode gap). The beam movement was enabled by removing the oxide layer underneath the beam by etching with hydrofloric acid.
The details of the experiment can be found in Melnikov \textit{et al.} \cite{Melnikov.2021}.
Furthermore, a small compression stress of \unit[2.6]{MPa} was used for the zero-mode approximation.
The experimental findings are well reproduced by the simple LPM developed in this paper, as can be seen in Fig.~\ref{fig:onions}c.

In summary we find that the zero-mode approximation gives rise to a simple LPM with a single degree of freedom, well suited to quantitatively describe all stable and unstable equilibria of clamped-clamped Coulomb actuated prismatic Euler-Bernoulli beams. 

%
\section{Discussion}

Spitz \textit{et al.} \cite{Spitz.2019} observed that the performance of a fairly complex MEMS µSpeaker can be successfully modelled by a heuristic single degree of freedom lumped parameter model.
Motivated by this research Melnikov \textit{et al.} \cite{Melnikov.2021} revisited the analysis of the bending profile of a Coulomb-activated prismatic micro beam, clamped at both ends: The study clearly confirms that the bending profile stays almost identical to the shape of the Euler-Bernoulli zero mode, independent of the load. This is true for the entire applicable voltage range within a very small error margin. The observations of Melnikov \textit{et al.} \cite{Melnikov.2021} allowed us here to develop the single degree of freedom lumped parameter model Eq.~\eqref{eq:LPM}, capable of accurately describing all stable and unstable equilibria of this highly non-linear electro-mechanical system. 
To the best of our knowledge, the existence of an accurate single degree of freedom LPM is not reported in the literature. In fact literature claims  the need for higher modes.\cite{M.I.Younis.2003}


The zero-mode approximation requires a method correctly projecting the Coulomb force onto the one dimensional Hilbert subspace, spanned by the Euler-Bernoulli zero-mode Eq.~\eqref{eq:zero_mode_ansatz}. Such projection is a global task in function space, that can not be performed using local techniques, such as a plain Taylor expansion. The ideas of Chebyshev and Edgeworth, underlying the original proof of the celebrated central limit theorem, furnish us here with the required means. As a result we obtain the analytical projection formula Eq.~\eqref{eq:f0_Li} for the Coulomb force. This formula allows us to extract the exact form of the contact singularity of the projected Coulomb force. Based on this knowledge, a global analysis of the Coulomb integral can be performed, leading to the handy algebraic expression  Eq.~\eqref{eq:f0_hat}. This completes the derivation of our highly accurate and simple to use lumped parameter model. To the best of our knowledge this is the first time, that Chebyshev-Edgeworth methods have been successfully used to solve a non-linear differential equation.

The results presented above now allow to efficiently compute the detailed frequency response and harmonic distortion of electromechanic MEMS transducers, with little computational effort. For practical applications, such dynamic computations are enabled by the large spectral distance of the Euler-Bernoulli zero-mode from higher Euler-Bernoulli modes, see Melnikov \textit{et al.} \cite{Melnikov.2021}.
\textcolor{black}{We note that the approach is applicable not only to clamped-clamped microbeams, but also to other conditions such as pinned-pinned or clamped-free.
In such a case, Eq.~\eqref{eq:zero_mode_ansatz} can stay the same while Eq.~\eqref{eq:beta_definition} changes, resulting in a new beta and new coefficient in Eq.~\eqref{eq:zero_mode_ansatz}.}
Certainly, time dependent FEM simulations will always allow to handle substantially more complex MEMS actuator geometries.
However the process of basic actuator design, as well as the circuit simulation of complex systems embracing MEMS actuators, see Monsalve \textit{et al.} \cite{Monsalve.2021}, 
greatly benefit from the availability of powerful LPM models.

We have presented the use of the Chebyshev-Edgeworth methods in this publications to model a very particular situation. While our focus on a simple case may help to understand the basic principle, it probably is misleading at the same time. Chebyshev-Edgeworth methods apply to far more general situations and allow for a broad range of applications. These include different boundary conditions, non-prismatic beams, the modelling of squeeze film damping, the computation of electric fringe field corrections and of contact forces to name a few. For the sake of clarity, we defer sharing the details of such generalizations to forthcoming publications.
\section{Conclusion}
All stable and unstable equilibrium states of Coulomb actuated prismatic clamped-clamped Euler-Bernoulli beams can be accurately computed by the simple to use lumped parameter model Eq.~\eqref{eq:LPM}. This LPM features only one degree of freedom, i.e., the amplitude of the Euler-Bernoulli zero-mode.
The contradiction of our results with previous findings of other groups are easily understood in terms of the advanced methods outlined above to adequately treat the Coulomb singularity.

\textcolor{black}{
The idea of the Chebyshev-Edgeworth expansion for the solution of nonlinear partial differential equations, which originates from probability theory, is not limited to beam mechanics.
We believe that our approach enables new insights into the derivation of highly effective lumped parameter models in a wide range of applications beyond elasticity theory.}


\begin{thebibliography}{33}%
\makeatletter
\providecommand \@ifxundefined [1]{%
 \@ifx{#1\undefined}
}%
\providecommand \@ifnum [1]{%
 \ifnum #1\expandafter \@firstoftwo
 \else \expandafter \@secondoftwo
 \fi
}%
\providecommand \@ifx [1]{%
 \ifx #1\expandafter \@firstoftwo
 \else \expandafter \@secondoftwo
 \fi
}%
\providecommand \natexlab [1]{#1}%
\providecommand \enquote  [1]{``#1''}%
\providecommand \bibnamefont  [1]{#1}%
\providecommand \bibfnamefont [1]{#1}%
\providecommand \citenamefont [1]{#1}%
\providecommand \href@noop [0]{\@secondoftwo}%
\providecommand \href [0]{\begingroup \@sanitize@url \@href}%
\providecommand \@href[1]{\@@startlink{#1}\@@href}%
\providecommand \@@href[1]{\endgroup#1\@@endlink}%
\providecommand \@sanitize@url [0]{\catcode `\\12\catcode `\$12\catcode
  `\&12\catcode `\#12\catcode `\^12\catcode `\_12\catcode `\%12\relax}%
\providecommand \@@startlink[1]{}%
\providecommand \@@endlink[0]{}%
\providecommand \url  [0]{\begingroup\@sanitize@url \@url }%
\providecommand \@url [1]{\endgroup\@href {#1}{\urlprefix }}%
\providecommand \urlprefix  [0]{URL }%
\providecommand \Eprint [0]{\href }%
\providecommand \doibase [0]{https://doi.org/}%
\providecommand \selectlanguage [0]{\@gobble}%
\providecommand \bibinfo  [0]{\@secondoftwo}%
\providecommand \bibfield  [0]{\@secondoftwo}%
\providecommand \translation [1]{[#1]}%
\providecommand \BibitemOpen [0]{}%
\providecommand \bibitemStop [0]{}%
\providecommand \bibitemNoStop [0]{.\EOS\space}%
\providecommand \EOS [0]{\spacefactor3000\relax}%
\providecommand \BibitemShut  [1]{\csname bibitem#1\endcsname}%
\let\auto@bib@innerbib\@empty
\bibitem [{\citenamefont {Senturia}(2002)}]{Senturia.2002}%
  \BibitemOpen
  \bibfield  {author} {\bibinfo {author} {\bibfnamefont {S.~D.}\ \bibnamefont
  {Senturia}},\ }\href {https://doi.org/10.1007/b117574} {\emph {\bibinfo
  {title} {{Microsystem Design}}}}\ (\bibinfo  {publisher} {{Kluwer Academic
  Publishers}},\ \bibinfo {address} {Boston, MA},\ \bibinfo {year}
  {2002})\BibitemShut {NoStop}%
\bibitem [{\citenamefont {Leondes}(2006)}]{Leondes.2006}%
  \BibitemOpen
  \bibfield  {author} {\bibinfo {author} {\bibfnamefont {C.~T.}\ \bibnamefont
  {Leondes}},\ }\href {https://doi.org/10.1007/b136111} {\emph {\bibinfo
  {title} {{MEMS/NEMS: Handbook Techniques and Applications}}}}\ (\bibinfo
  {publisher} {{Springer Science+Business Media Inc}},\ \bibinfo {address}
  {Boston, MA},\ \bibinfo {year} {2006})\BibitemShut {NoStop}%
\bibitem [{\citenamefont {Hsu}(2008)}]{Hsu.2008}%
  \BibitemOpen
  \bibfield  {author} {\bibinfo {author} {\bibfnamefont {T.-R.}\ \bibnamefont
  {Hsu}},\ }\href
  {http://www.loc.gov/catdir/enhancements/fy0741/2007017041-d.html} {\emph
  {\bibinfo {title} {{MEMS and microsystems: Design, manufacture, and nanoscale
  engineering}}}},\ \bibinfo {edition} {2nd}\ ed.\ (\bibinfo  {publisher}
  {Wiley},\ \bibinfo {address} {Hoboken, NJ},\ \bibinfo {year}
  {2008})\BibitemShut {NoStop}%
\bibitem [{\citenamefont {Saliterman}(2006)}]{Saliterman.2006}%
  \BibitemOpen
  \bibfield  {author} {\bibinfo {author} {\bibfnamefont {S.~S.}\ \bibnamefont
  {Saliterman}},\ }\href
  {http://www.loc.gov/catdir/enhancements/fy0659/2005015759-d.html} {\emph
  {\bibinfo {title} {{Fundamentals of BioMEMS and medical microdevices}}}},\
  \bibinfo {series} {{SPIE Press monograph series}}, Vol.\ \bibinfo {volume}
  {153}\ (\bibinfo  {publisher} {{SPIE Press}},\ \bibinfo {address}
  {Bellingham, WA},\ \bibinfo {year} {2006})\BibitemShut {NoStop}%
\bibitem [{\citenamefont {Lucyszyn}(2004)}]{Lucyszyn.2004}%
  \BibitemOpen
  \bibfield  {author} {\bibinfo {author} {\bibfnamefont {S.}~\bibnamefont
  {Lucyszyn}},\ }\bibfield  {title} {\enquote {\bibinfo {title} {{Review of
  radio frequency microelectromechanical systems technology}},}\ }\href
  {https://doi.org/10.1049/ip-smt:20040405} {\bibfield  {journal} {\bibinfo
  {journal} {{IEE Proceedings - Science, Measurement and Technology}}\ }\textbf
  {\bibinfo {volume} {151}},\ \bibinfo {pages} {93--103} (\bibinfo {year}
  {2004})}\BibitemShut {NoStop}%
\bibitem [{\citenamefont {Li}, \citenamefont {Xu},\ and\ \citenamefont
  {Zhao}(2018)}]{Li.2018}%
  \BibitemOpen
  \bibfield  {author} {\bibinfo {author} {\bibfnamefont {S.}~\bibnamefont
  {Li}}, \bibinfo {author} {\bibfnamefont {L.~D.}\ \bibnamefont {Xu}},\ and\
  \bibinfo {author} {\bibfnamefont {S.}~\bibnamefont {Zhao}},\ }\bibfield
  {title} {\enquote {\bibinfo {title} {{5G Internet of Things: A survey}},}\
  }\href {https://doi.org/10.1016/j.jii.2018.01.005} {\bibfield  {journal}
  {\bibinfo  {journal} {{Journal of Industrial Information Integration}}\
  }\textbf {\bibinfo {volume} {10}},\ \bibinfo {pages} {1--9} (\bibinfo {year}
  {2018})}\BibitemShut {NoStop}%
\bibitem [{\citenamefont {Coppa}\ \emph {et~al.}(2007)\citenamefont {Coppa},
  \citenamefont {Cianci}, \citenamefont {Foglietti}, \citenamefont {Caliano},\
  and\ \citenamefont {Pappalardo}}]{Coppa.2007}%
  \BibitemOpen
  \bibfield  {author} {\bibinfo {author} {\bibfnamefont {A.}~\bibnamefont
  {Coppa}}, \bibinfo {author} {\bibfnamefont {E.}~\bibnamefont {Cianci}},
  \bibinfo {author} {\bibfnamefont {V.}~\bibnamefont {Foglietti}}, \bibinfo
  {author} {\bibfnamefont {G.}~\bibnamefont {Caliano}},\ and\ \bibinfo {author}
  {\bibfnamefont {M.}~\bibnamefont {Pappalardo}},\ }\bibfield  {title}
  {\enquote {\bibinfo {title} {{Building CMUTs for imaging applications from
  top to bottom}},}\ }\href {https://doi.org/10.1016/j.mee.2007.01.211}
  {\bibfield  {journal} {\bibinfo  {journal} {{Microelectronic Engineering}}\
  }\textbf {\bibinfo {volume} {84}},\ \bibinfo {pages} {1312--1315} (\bibinfo
  {year} {2007})}\BibitemShut {NoStop}%
\bibitem [{\citenamefont {{S. Finkbeiner}}(2013)}]{S.Finkbeiner.2013}%
  \BibitemOpen
  \bibfield  {author} {\bibinfo {author} {\bibnamefont {{S. Finkbeiner}}},\
  }\bibfield  {title} {\enquote {\bibinfo {title} {{MEMS for automotive and
  consumer electronics}},}\ }in\ \href
  {https://doi.org/10.1109/ESSCIRC.2013.6649059} {\emph {\bibinfo {booktitle}
  {{2013 Proceedings of the ESSCIRC (ESSCIRC)}}}}\ (\bibinfo {year} {2013})\
  pp.\ \bibinfo {pages} {9--14}\BibitemShut {NoStop}%
\bibitem [{\citenamefont {Zou}, \citenamefont {Thiruvenkatanathan},\ and\
  \citenamefont {Seshia}(2014)}]{Zou.2014}%
  \BibitemOpen
  \bibfield  {author} {\bibinfo {author} {\bibfnamefont {X.}~\bibnamefont
  {Zou}}, \bibinfo {author} {\bibfnamefont {P.}~\bibnamefont
  {Thiruvenkatanathan}},\ and\ \bibinfo {author} {\bibfnamefont {A.~A.}\
  \bibnamefont {Seshia}},\ }\bibfield  {title} {\enquote {\bibinfo {title} {{A
  Seismic-Grade Resonant MEMS Accelerometer}},}\ }\href
  {https://doi.org/10.1109/JMEMS.2014.2319196} {\bibfield  {journal} {\bibinfo
  {journal} {{Journal of Microelectromechanical Systems}}\ }\textbf {\bibinfo
  {volume} {23}},\ \bibinfo {pages} {768--770} (\bibinfo {year}
  {2014})}\BibitemShut {NoStop}%
\bibitem [{\citenamefont {Verdot}\ \emph {et~al.}(2016)\citenamefont {Verdot},
  \citenamefont {Redon}, \citenamefont {Ege}, \citenamefont {Czarny},
  \citenamefont {Guianvarc'h},\ and\ \citenamefont {Guyader}}]{Verdot.2016}%
  \BibitemOpen
  \bibfield  {author} {\bibinfo {author} {\bibfnamefont {T.}~\bibnamefont
  {Verdot}}, \bibinfo {author} {\bibfnamefont {E.}~\bibnamefont {Redon}},
  \bibinfo {author} {\bibfnamefont {K.}~\bibnamefont {Ege}}, \bibinfo {author}
  {\bibfnamefont {J.}~\bibnamefont {Czarny}}, \bibinfo {author} {\bibfnamefont
  {C.}~\bibnamefont {Guianvarc'h}},\ and\ \bibinfo {author} {\bibfnamefont
  {J.-L.}\ \bibnamefont {Guyader}},\ }\bibfield  {title} {\enquote {\bibinfo
  {title} {{Microphone with Planar Nano-Gauge Detection: Fluid-Structure
  Coupling Including Thermoviscous Effects}},}\ }\href
  {https://doi.org/10.3813/AAA.918969} {\bibfield  {journal} {\bibinfo
  {journal} {{Acta Acustica united with Acustica}}\ }\textbf {\bibinfo {volume}
  {102}},\ \bibinfo {pages} {517--529} (\bibinfo {year} {2016})}\BibitemShut
  {NoStop}%
\bibitem [{\citenamefont {Shahosseini}\ \emph {et~al.}(2013)\citenamefont
  {Shahosseini}, \citenamefont {Lefeuvre}, \citenamefont {Moulin},
  \citenamefont {Martincic}, \citenamefont {Woytasik},\ and\ \citenamefont
  {Lemarquand}}]{Shahosseini.2013}%
  \BibitemOpen
  \bibfield  {author} {\bibinfo {author} {\bibfnamefont {I.}~\bibnamefont
  {Shahosseini}}, \bibinfo {author} {\bibfnamefont {E.}~\bibnamefont
  {Lefeuvre}}, \bibinfo {author} {\bibfnamefont {J.}~\bibnamefont {Moulin}},
  \bibinfo {author} {\bibfnamefont {E.}~\bibnamefont {Martincic}}, \bibinfo
  {author} {\bibfnamefont {M.}~\bibnamefont {Woytasik}},\ and\ \bibinfo
  {author} {\bibfnamefont {G.}~\bibnamefont {Lemarquand}},\ }\bibfield  {title}
  {\enquote {\bibinfo {title} {{Optimization and Microfabrication of High
  Performance Silicon-Based MEMS Microspeaker}},}\ }\href
  {https://doi.org/10.1109/JSEN.2012.2213807} {\bibfield  {journal} {\bibinfo
  {journal} {{IEEE Sensors Journal}}\ }\textbf {\bibinfo {volume} {13}},\
  \bibinfo {pages} {273--284} (\bibinfo {year} {2013})}\BibitemShut {NoStop}%
\bibitem [{\citenamefont {Kaiser}\ \emph {et~al.}()\citenamefont {Kaiser},
  \citenamefont {Langa}, \citenamefont {Ehrig}, \citenamefont {Stolz},
  \citenamefont {Schenk}, \citenamefont {Conrad}, \citenamefont {Schenk},
  \citenamefont {Schimmanz},\ and\ \citenamefont {Schuffenhauer}}]{Kaiser.}%
  \BibitemOpen
  \bibfield  {author} {\bibinfo {author} {\bibfnamefont {B.}~\bibnamefont
  {Kaiser}}, \bibinfo {author} {\bibfnamefont {S.}~\bibnamefont {Langa}},
  \bibinfo {author} {\bibfnamefont {L.}~\bibnamefont {Ehrig}}, \bibinfo
  {author} {\bibfnamefont {M.}~\bibnamefont {Stolz}}, \bibinfo {author}
  {\bibfnamefont {H.}~\bibnamefont {Schenk}}, \bibinfo {author} {\bibfnamefont
  {H.}~\bibnamefont {Conrad}}, \bibinfo {author} {\bibfnamefont
  {H.}~\bibnamefont {Schenk}}, \bibinfo {author} {\bibfnamefont
  {K.}~\bibnamefont {Schimmanz}},\ and\ \bibinfo {author} {\bibfnamefont
  {D.}~\bibnamefont {Schuffenhauer}},\ }\bibfield  {title} {\enquote {\bibinfo
  {title} {{Concept and proof for an all-silicon MEMS micro speaker utilizing
  air chambers}},}\ }\href {https://doi.org/10.1038/s41378-019-0095-9}
  {\bibfield  {journal} {\bibinfo  {journal} {{Microsystems {\&}
  Nanoengineering}}\ }\textbf {\bibinfo {volume} {5}},\ \bibinfo {pages}
  {1--11}}\BibitemShut {NoStop}%
\bibitem [{\citenamefont {Kim}, \citenamefont {Thang},\ and\ \citenamefont
  {Kim}(2009)}]{Kim.2009}%
  \BibitemOpen
  \bibfield  {author} {\bibinfo {author} {\bibfnamefont {J.-H.}\ \bibnamefont
  {Kim}}, \bibinfo {author} {\bibfnamefont {N.~D.}\ \bibnamefont {Thang}},\
  and\ \bibinfo {author} {\bibfnamefont {T.-S.}\ \bibnamefont {Kim}},\
  }\bibfield  {title} {\enquote {\bibinfo {title} {{3-D hand motion tracking
  and gesture recognition using a data glove}},}\ }in\ \href
  {https://doi.org/10.1109/ISIE.2009.5221998} {\emph {\bibinfo {booktitle}
  {{IEEE International Symposium on Industrial Electronics, 2009}}}}\ (\bibinfo
   {publisher} {IEEE},\ \bibinfo {address} {Piscataway, NJ},\ \bibinfo {year}
  {2009})\ pp.\ \bibinfo {pages} {1013--1018}\BibitemShut {NoStop}%
\bibitem [{\citenamefont {Bianzino}\ \emph {et~al.}()\citenamefont {Bianzino},
  \citenamefont {Chaudet}, \citenamefont {Rossi},\ and\ \citenamefont
  {Rougier}}]{Bianzino.2010}%
  \BibitemOpen
  \bibfield  {author} {\bibinfo {author} {\bibfnamefont {A.~P.}\ \bibnamefont
  {Bianzino}}, \bibinfo {author} {\bibfnamefont {C.}~\bibnamefont {Chaudet}},
  \bibinfo {author} {\bibfnamefont {D.}~\bibnamefont {Rossi}},\ and\ \bibinfo
  {author} {\bibfnamefont {J.-L.}\ \bibnamefont {Rougier}},\ }\href
  {https://arxiv.org/pdf/1010.3880} {\enquote {\bibinfo {title} {{A Survey of
  Green Networking Research}},}\ }\BibitemShut {NoStop}%
\bibitem [{\citenamefont {Worthington}(2017)}]{Worthington.2017}%
  \BibitemOpen
  \bibfield  {author} {\bibinfo {author} {\bibfnamefont {T.}~\bibnamefont
  {Worthington}},\ }\href@noop {} {\emph {\bibinfo {title} {{ICT
  Sustainability: Assessment and strategies for a low carbon future}}}}\
  (\bibinfo  {publisher} {{LULU COM}},\ \bibinfo {address} {[Place of
  publication not identified]},\ \bibinfo {year} {2017})\BibitemShut {NoStop}%
\bibitem [{\citenamefont {Melnikov}\ \emph {et~al.}(2021)\citenamefont
  {Melnikov}, \citenamefont {Schenk}, \citenamefont {Monsalve}, \citenamefont
  {Wall}, \citenamefont {Stolz}, \citenamefont {Mrosk}, \citenamefont {Langa},\
  and\ \citenamefont {Kaiser}}]{Melnikov.2021}%
  \BibitemOpen
  \bibfield  {author} {\bibinfo {author} {\bibfnamefont {A.}~\bibnamefont
  {Melnikov}}, \bibinfo {author} {\bibfnamefont {H.~A.~G.}\ \bibnamefont
  {Schenk}}, \bibinfo {author} {\bibfnamefont {J.~M.}\ \bibnamefont
  {Monsalve}}, \bibinfo {author} {\bibfnamefont {F.}~\bibnamefont {Wall}},
  \bibinfo {author} {\bibfnamefont {M.}~\bibnamefont {Stolz}}, \bibinfo
  {author} {\bibfnamefont {A.}~\bibnamefont {Mrosk}}, \bibinfo {author}
  {\bibfnamefont {S.}~\bibnamefont {Langa}},\ and\ \bibinfo {author}
  {\bibfnamefont {B.}~\bibnamefont {Kaiser}},\ }\bibfield  {title} {\enquote
  {\bibinfo {title} {{Coulomb-actuated microbeams revisited: experimental and
  numerical modal decomposition of the saddle-node bifurcation}},}\ }\href
  {https://doi.org/10.1038/s41378-021-00265-y} {\bibfield  {journal} {\bibinfo
  {journal} {{Microsystems {\&} Nanoengineering}}\ }\textbf {\bibinfo {volume}
  {7}} (\bibinfo {year} {2021}),\ 10.1038/s41378-021-00265-y}\BibitemShut
  {NoStop}%
\bibitem [{\citenamefont {Younis}, \citenamefont {Abdel-Rahman},\ and\
  \citenamefont {Nayfeh}(2002)}]{Younis.2002}%
  \BibitemOpen
  \bibfield  {author} {\bibinfo {author} {\bibfnamefont {M.}~\bibnamefont
  {Younis}}, \bibinfo {author} {\bibfnamefont {E.}~\bibnamefont
  {Abdel-Rahman}},\ and\ \bibinfo {author} {\bibfnamefont {A.}~\bibnamefont
  {Nayfeh}},\ }\bibfield  {title} {\enquote {\bibinfo {title} {{Static and
  Dynamic Behavior of an Electrically Excited Resonant Microbeam}},}\ }in\
  \href {https://doi.org/10.2514/6.2002-1305} {\emph {\bibinfo {booktitle}
  {{Structures, Structural Dynamics, and Materials and Co-located
  Conferences}}}}\ (\bibinfo  {publisher} {{[publisher not identified]}},\
  \bibinfo {address} {[Place of publication not identified]},\ \bibinfo {year}
  {2002})\BibitemShut {NoStop}%
\bibitem [{\citenamefont {{Eihab M Abdel-Rahman}}, \citenamefont {{Mohammad I
  Younis}},\ and\ \citenamefont {{Ali H
  Nayfeh}}(2002)}]{EihabMAbdelRahman.2002}%
  \BibitemOpen
  \bibfield  {author} {\bibinfo {author} {\bibnamefont {{Eihab M
  Abdel-Rahman}}}, \bibinfo {author} {\bibnamefont {{Mohammad I Younis}}},\
  and\ \bibinfo {author} {\bibnamefont {{Ali H Nayfeh}}},\ }\bibfield  {title}
  {\enquote {\bibinfo {title} {{Characterization of the mechanical behavior of
  an electrically actuated microbeam}},}\ }\href
  {https://doi.org/10.1088/0960-1317/12/6/306} {\bibfield  {journal} {\bibinfo
  {journal} {{Journal of Micromechanics and Microengineering}}\ }\textbf
  {\bibinfo {volume} {12}},\ \bibinfo {pages} {759} (\bibinfo {year}
  {2002})}\BibitemShut {NoStop}%
\bibitem [{\citenamefont {{M. I. Younis}}, \citenamefont {{E. M.
  Abdel-Rahman}},\ and\ \citenamefont {{A. Nayfeh}}(2003)}]{M.I.Younis.2003}%
  \BibitemOpen
  \bibfield  {author} {\bibinfo {author} {\bibnamefont {{M. I. Younis}}},
  \bibinfo {author} {\bibnamefont {{E. M. Abdel-Rahman}}},\ and\ \bibinfo
  {author} {\bibnamefont {{A. Nayfeh}}},\ }\bibfield  {title} {\enquote
  {\bibinfo {title} {{A reduced-order model for electrically actuated
  microbeam-based MEMS}},}\ }\href {https://doi.org/10.1109/JMEMS.2003.818069}
  {\bibfield  {journal} {\bibinfo  {journal} {{Journal of
  Microelectromechanical Systems}}\ }\textbf {\bibinfo {volume} {12}},\
  \bibinfo {pages} {672--680} (\bibinfo {year} {2003})}\BibitemShut {NoStop}%
\bibitem [{\citenamefont {{M. I. Younis}}\ and\ \citenamefont {{A. H.
  Nayfeh}}(2003)}]{M.I.Younis.2003b}%
  \BibitemOpen
  \bibfield  {author} {\bibinfo {author} {\bibnamefont {{M. I. Younis}}}\ and\
  \bibinfo {author} {\bibnamefont {{A. H. Nayfeh}}},\ }\bibfield  {title}
  {\enquote {\bibinfo {title} {{A Study of the Nonlinear Response of a Resonant
  Microbeam to an Electric Actuation}},}\ }\href
  {https://doi.org/10.1023/A:1022103118330} {\bibfield  {journal} {\bibinfo
  {journal} {{Nonlinear Dynamics}}\ }\textbf {\bibinfo {volume} {31}},\
  \bibinfo {pages} {91--117} (\bibinfo {year} {2003})}\BibitemShut {NoStop}%
\bibitem [{\citenamefont {Nayfeh}, \citenamefont {Younis},\ and\ \citenamefont
  {Abdel-Rahman}(2005)}]{Nayfeh.2005}%
  \BibitemOpen
  \bibfield  {author} {\bibinfo {author} {\bibfnamefont {A.~H.}\ \bibnamefont
  {Nayfeh}}, \bibinfo {author} {\bibfnamefont {M.~I.}\ \bibnamefont {Younis}},\
  and\ \bibinfo {author} {\bibfnamefont {E.~M.}\ \bibnamefont {Abdel-Rahman}},\
  }\bibfield  {title} {\enquote {\bibinfo {title} {{Reduced-Order Models for
  MEMS Applications}},}\ }\href {https://doi.org/10.1007/s11071-005-2809-9}
  {\bibfield  {journal} {\bibinfo  {journal} {{Nonlinear Dynamics}}\ }\textbf
  {\bibinfo {volume} {41}},\ \bibinfo {pages} {211--236} (\bibinfo {year}
  {2005})}\BibitemShut {NoStop}%
\bibitem [{\citenamefont {Nayfeh}, \citenamefont {Younis},\ and\ \citenamefont
  {Abdel-Rahman}(2007)}]{Nayfeh.2007}%
  \BibitemOpen
  \bibfield  {author} {\bibinfo {author} {\bibfnamefont {A.~H.}\ \bibnamefont
  {Nayfeh}}, \bibinfo {author} {\bibfnamefont {M.~I.}\ \bibnamefont {Younis}},\
  and\ \bibinfo {author} {\bibfnamefont {E.~M.}\ \bibnamefont {Abdel-Rahman}},\
  }\bibfield  {title} {\enquote {\bibinfo {title} {{Dynamic pull-in phenomenon
  in MEMS resonators}},}\ }\href {https://doi.org/10.1007/s11071-006-9079-z}
  {\bibfield  {journal} {\bibinfo  {journal} {{Nonlinear Dynamics}}\ }\textbf
  {\bibinfo {volume} {48}},\ \bibinfo {pages} {153--163} (\bibinfo {year}
  {2007})}\BibitemShut {NoStop}%
\bibitem [{\citenamefont {Younis}(2011)}]{Younis.2011}%
  \BibitemOpen
  \bibfield  {author} {\bibinfo {author} {\bibfnamefont {M.~I.}\ \bibnamefont
  {Younis}},\ }\href {https://doi.org/10.1007/978-1-4419-6020-7} {\emph
  {\bibinfo {title} {{MEMS Linear and Nonlinear Statics and Dynamics}}}},\
  \bibinfo {series} {{Microsystems}}, Vol.~\bibinfo {volume} {20}\ (\bibinfo
  {publisher} {{Springer Science+Business Media LLC}},\ \bibinfo {address}
  {Boston, MA},\ \bibinfo {year} {2011})\BibitemShut {NoStop}%
\bibitem [{\citenamefont {Tchebycheff}(1890)}]{Tchebycheff.1890}%
  \BibitemOpen
  \bibfield  {author} {\bibinfo {author} {\bibfnamefont {P.}~\bibnamefont
  {Tchebycheff}},\ }\bibfield  {title} {\enquote {\bibinfo {title} {{Sur deux
  th{\'e}or{\`e}mes relatifs aux probabilit{\'e}s}},}\ }\href@noop {}
  {\bibfield  {journal} {\bibinfo  {journal} {{Acta Mathematica}}\ ,\ \bibinfo
  {pages} {305--315}} (\bibinfo {year} {1890})}\BibitemShut {NoStop}%
\bibitem [{\citenamefont {Edgeworth}(1905)}]{Edgeworth.1905}%
  \BibitemOpen
  \bibfield  {author} {\bibinfo {author} {\bibfnamefont {F.~Y.}\ \bibnamefont
  {Edgeworth}},\ }\bibfield  {title} {\enquote {\bibinfo {title} {{The Law of
  Error}},}\ }\href@noop {} {\bibfield  {journal} {\bibinfo  {journal}
  {{Cambridge Philos. Soc.}}\ ,\ \bibinfo {pages} {36--66, 113--141}} (\bibinfo
  {year} {1905})}\BibitemShut {NoStop}%
\bibitem [{\citenamefont {Forster}(2017)}]{Forster.2017}%
  \BibitemOpen
  \bibfield  {author} {\bibinfo {author} {\bibfnamefont {O.}~\bibnamefont
  {Forster}},\ }\href {https://doi.org/10.1007/978-3-658-16746-2} {\emph
  {\bibinfo {title} {{Analysis 3: Ma{\ss}- und Integrationstheorie,
  Integrals{\"a}tze im IRn und Anwendungen}}}},\ \bibinfo {edition} {8th}\
  ed.,\ {Aufbaukurs Mathematik}\ (\bibinfo  {publisher} {{Springer Spektrum}},\
  \bibinfo {address} {Wiesbaden},\ \bibinfo {year} {2017})\BibitemShut
  {NoStop}%
\bibitem [{\citenamefont {Wallace}(1958)}]{Wallace.1958}%
  \BibitemOpen
  \bibfield  {author} {\bibinfo {author} {\bibfnamefont {D.~L.}\ \bibnamefont
  {Wallace}},\ }\bibfield  {title} {\enquote {\bibinfo {title} {{Asymptotic
  Approximations to Distributions}},}\ }\href@noop {} {\bibfield  {journal}
  {\bibinfo  {journal} {{Ann. Math. Stat.}}\ ,\ \bibinfo {pages} {635--654}}
  (\bibinfo {year} {1958})}\BibitemShut {NoStop}%
\bibitem [{\citenamefont {K{\"o}nigsberger}(2001)}]{Konigsberger.2001}%
  \BibitemOpen
  \bibfield  {author} {\bibinfo {author} {\bibfnamefont {K.}~\bibnamefont
  {K{\"o}nigsberger}},\ }\href {https://doi.org/10.1007/978-3-642-97890-6}
  {\emph {\bibinfo {title} {{Analysis 1}}}},\ {Springer-Lehrbuch}\ (\bibinfo
  {publisher} {{Springer Berlin Heidelberg}},\ \bibinfo {address} {Berlin,
  Heidelberg},\ \bibinfo {year} {2001})\BibitemShut {NoStop}%
\bibitem [{\citenamefont {Jonqui{\`e}re}(1889)}]{Jonquiere.1889}%
  \BibitemOpen
  \bibfield  {author} {\bibinfo {author} {\bibfnamefont {A.}~\bibnamefont
  {Jonqui{\`e}re}},\ }\bibfield  {title} {\enquote {\bibinfo {title} {{Note sur
  la s{\'e}rie {\$}sum{\_}(n=1){\^{}}(n=infty)(x{\^{}}n)/(n{\^{}}s){\$}.}}}\
  }\href@noop {} {\bibfield  {journal} {\bibinfo  {journal} {{Bull. Soc. Math.
  France}}\ ,\ \bibinfo {pages} {142--152}} (\bibinfo {year}
  {1889})}\BibitemShut {NoStop}%
\bibitem [{\citenamefont {Riemann}(1859)}]{Riemann.1859}%
  \BibitemOpen
  \bibfield  {author} {\bibinfo {author} {\bibfnamefont {B.}~\bibnamefont
  {Riemann}},\ }\bibfield  {title} {\enquote {\bibinfo {title} {{{\"U}ber die
  Anzahl der Primzahlen unter einer gegebenen Gr{\"o}sse}},}\ }\href@noop {}
  {\bibfield  {journal} {\bibinfo  {journal} {{Monatsber. K{\"o}nigl. Preuss.
  Akad. Wiss. Berlin}}\ ,\ \bibinfo {pages} {671--680}} (\bibinfo {year}
  {1859})}\BibitemShut {NoStop}%
\bibitem [{\citenamefont {Gilbert}, \citenamefont {Ananthasuresh},\ and\
  \citenamefont {Senturia}(1996)}]{Gilbert.1996}%
  \BibitemOpen
  \bibfield  {author} {\bibinfo {author} {\bibfnamefont {J.~R.}\ \bibnamefont
  {Gilbert}}, \bibinfo {author} {\bibfnamefont {G.~K.}\ \bibnamefont
  {Ananthasuresh}},\ and\ \bibinfo {author} {\bibfnamefont {S.~D.}\
  \bibnamefont {Senturia}},\ }\bibfield  {title} {\enquote {\bibinfo {title}
  {{3D modeling of contact problems and hysteresis in coupled
  electro-mechanics}},}\ }in\ \href
  {https://doi.org/10.1109/MEMSYS.1996.493841} {\emph {\bibinfo {booktitle}
  {{Proceedings / IEEE the Ninth Annual International Workshop on Micro Electro
  Mechanical Systems [MEMS]}}}}\ (\bibinfo  {publisher} {{IEEE Service
  Center}},\ \bibinfo {address} {Piscataway, NJ},\ \bibinfo {year} {1996})\
  pp.\ \bibinfo {pages} {127--132}\BibitemShut {NoStop}%
\bibitem [{\citenamefont {Spitz}\ \emph {et~al.}(2019)\citenamefont {Spitz},
  \citenamefont {Wall}, \citenamefont {Schenk}, \citenamefont {Melnikov},
  \citenamefont {Ehrig}, \citenamefont {Langa}, \citenamefont {Stolz},
  \citenamefont {Kaiser}, \citenamefont {Conrad}, \citenamefont {Schenk},\ and\
  \citenamefont {Pufe}}]{Spitz.2019}%
  \BibitemOpen
  \bibfield  {author} {\bibinfo {author} {\bibfnamefont {B.}~\bibnamefont
  {Spitz}}, \bibinfo {author} {\bibfnamefont {F.}~\bibnamefont {Wall}},
  \bibinfo {author} {\bibfnamefont {H.}~\bibnamefont {Schenk}}, \bibinfo
  {author} {\bibfnamefont {A.}~\bibnamefont {Melnikov}}, \bibinfo {author}
  {\bibfnamefont {L.}~\bibnamefont {Ehrig}}, \bibinfo {author} {\bibfnamefont
  {S.}~\bibnamefont {Langa}}, \bibinfo {author} {\bibfnamefont
  {M.}~\bibnamefont {Stolz}}, \bibinfo {author} {\bibfnamefont
  {B.}~\bibnamefont {Kaiser}}, \bibinfo {author} {\bibfnamefont
  {H.}~\bibnamefont {Conrad}}, \bibinfo {author} {\bibfnamefont
  {H.}~\bibnamefont {Schenk}},\ and\ \bibinfo {author} {\bibfnamefont
  {W.}~\bibnamefont {Pufe}},\ }\bibfield  {title} {\enquote {\bibinfo {title}
  {{Audio-Transducer for In-Ear-Applications based on CMOS compatible
  electrostatic actuators}},}\ }\href@noop {} {\bibfield  {journal} {\bibinfo
  {journal} {{MikroSystemTechnik Kongress Proceedings}}\ } (\bibinfo {year}
  {2019})}\BibitemShut {NoStop}%
\bibitem [{\citenamefont {Monsalve}\ \emph {et~al.}(2021)\citenamefont
  {Monsalve}, \citenamefont {Melnikov}, \citenamefont {Kaiser}, \citenamefont
  {Schuffenhauer}, \citenamefont {Stolz}, \citenamefont {Ehrig}, \citenamefont
  {Schenk}, \citenamefont {Conrad},\ and\ \citenamefont
  {Schenk}}]{Monsalve.2021}%
  \BibitemOpen
  \bibfield  {author} {\bibinfo {author} {\bibfnamefont {J.~M.}\ \bibnamefont
  {Monsalve}}, \bibinfo {author} {\bibfnamefont {A.}~\bibnamefont {Melnikov}},
  \bibinfo {author} {\bibfnamefont {B.}~\bibnamefont {Kaiser}}, \bibinfo
  {author} {\bibfnamefont {D.}~\bibnamefont {Schuffenhauer}}, \bibinfo {author}
  {\bibfnamefont {M.}~\bibnamefont {Stolz}}, \bibinfo {author} {\bibfnamefont
  {L.}~\bibnamefont {Ehrig}}, \bibinfo {author} {\bibfnamefont {H.~A.}\
  \bibnamefont {Schenk}}, \bibinfo {author} {\bibfnamefont {H.}~\bibnamefont
  {Conrad}},\ and\ \bibinfo {author} {\bibfnamefont {H.}~\bibnamefont
  {Schenk}},\ }\bibfield  {title} {\enquote {\bibinfo {title} {Large-signal
  equivalent-circuit model of asymmetric electrostatic transducers},}\ }\href
  {https://doi.org/10.1109/TMECH.2021.3112267} {\bibfield  {journal} {\bibinfo
  {journal} {{IEEE/ASME Transactions on Mechatronics}}\ } (\bibinfo {year}
  {2021}),\ 10.1109/TMECH.2021.3112267}\BibitemShut {NoStop}%
\end{thebibliography}

%

\clearpage


\end{document}